# ATOMS: ALMA Three-millimeter Observations of Massive Star-forming regions -XIV. Properties of resolved UC HII regions

C. Zhang,[1,2] Feng-Yao Zhu,[3]⋆ Tie Liu,[4]† Z.-Y. Ren,[5] H.-L. Liu,[6] Ke Wang,[7] J.-W. Wu,[8,5] Y. Zhang,[9] J.-W. Zhou,[5] K. Tatematsu,[10] Guido Garay,[11] Anandmayee Tej,[12] Shanghuo Li,[13] W.F. Xu,[7] Chang Won Lee,[13] Leonardo Bronfman,[11] Archana Soam,[14] and D. Li[5,15]

[1] *Department of Physics, Taiyuan Normal University, Jinzhong 030619, China*
[2] *Institute of Computational and Applied Physics, Taiyuan Normal University, Jinzhong 030619, China*
[3] *Research Center for Intelligent Computing Platforms, Zhejiang Laboratory, Hangzhou, 311100, China*
[4] *Shanghai Astronomical Observatory, Chinese Academy of Sciences, 80 Nandan Road, Shanghai 200030, China*
[5] *National Astronomical Observatories, Chinese Academy of Sciences, Beijing, 100012, People's Republic of China*
[6] *Department of Astronomy, Yunnan University, Kunming, 650091, PR China*
[7] *Kavli Institute for Astronomy and Astrophysics, Peking University, 5 Yiheyuan Road, Haidian District, Beijing 100871, People's Republic of China*
[8] *University of Chinese Academy of Sciences, Beijing 100049, PR China*
[9] *School of Physics and Astronomy, Sun Yat-sen University, 2 Daxue Road, Zhuhai, Guangdong, 519082, People's Republic of China*
[10] *Nobeyama Radio Observatory, National Astronomical Observatory of Japan, National Institutes of Natural Sciences, Nobeyama, Minamimaki, Minamisaku, Nagano 384-*
[11] *Departamento de Astronoma, Universidad de Chile, Las Condes, Santiago, Chile*
[12] *Indian Institute of Space Science and Technology, Thiruvananthapuram 695 547, Kerala, India*
[13] *Korea Astronomy and Space Science Institute, 776 Daedeokdaero, Yuseong-gu, Daejeon 34055, Republic of Korea*
[14] *SOFIA Science Centre, USRA, NASA Ames Research Centre, MS-12, N232, Moffett Field, CA 94035, USA*
[15] *NAOC-UKZN Computational Astrophysics Centre, University of KwaZulu-Natal, Durban 4000, South Africa*

**ABSTRACT**

Hydrogen recombination lines (RRLs) are one of the major diagnostics of the physical properties of HII regions. We use RRL H40$\alpha$, He40$\alpha$ and 3 mm continuum emission to investigate the properties of a large sample of resolved UC HII regions identified in the ATOMS survey. In total, we identify 94 UC HII regions from H40$\alpha$ emission. The basic parameters for these UC HII regions such as electron density, emission measure, electron temperature, ionic abundance ratio ($n_{He^+}/n_{H^+}$), and line width are derived. The median electron density and the median $n_{He^+}/n_{H^+}$ ratio of these UC HII regions derived from RRLs are ∼9000 cm$^{-3}$ and 0.11, respectively. Within UC HII regions, the $n_{He^+}/n_{H^+}$ ratios derived from the intensity ratio of the He40$\alpha$ and H40$\alpha$ lines seems to be higher in the boundary region than in the center. The H40$\alpha$ line width is mainly broadened by thermal motion and microturbulence. The electron temperature of these UC HII regions has a median value of ∼6700 K, and its dependence on galactocentric distance is weak.

## 1 INTRODUCTION

Massive stars can significantly influence their local environments through powerful feedback such as winds, radiation and supernova explosions. Ionizing radiation from massive stars produces pronounced HII regions around them (e.g., Krumholz et al. 2014). Thus, observations of HII regions provide important insight into massive star formation (Habing & Israel 1979; Churchwell 2002; Zinnecker & Yorke 2007).

HII regions in the Galaxy show low $n_{He^+}/n_{H^+}$ number density ratios (less than 0.1 which is the actual He/H abundance; Draine 2011) from RRL observations, and this ratio seems to decrease with increasing distance from the HII region beyond the photodissociation region (PDR; Luisi et al. 2016; Roshi et al. 2017). Here $n_{He^+}$ is the number density of singly ionized helium (He$^+$), and $n_{H^+}$ is that of ionized hydrogen (H$^+$). The observed low $n_{He^+}/n_{H^+}$ number density ratios outside the PDR regions may indicate that there will be fewer photons with enough energy to ionize He compared to H there or part of radiation field is attenuated by interstellar dust (Roshi et al. 2012). Therefore, the $n_{He^+}/n_{H^+}$ ratio is very sensitive to the strength of radiation field and dust properties. The $n_{He^+}/n_{H^+}$ ratios inside UC HII regions, however, have not been systematically investigated.

Electron temperatures, T$_e$, of HII regions were estimated from the intensity ratios between radio recombination lines (RRLs) and continuum emission on the LTE approximation in previous observations (e.g. Shaver et al. 1983; Caswell & Haynes 1987). However, Gordon & Sorochenko (2002) points out that using LTE approxi-mation could cause significant deviation of estimated $T_e$ in some conditions. Previous studies indicate that $T_e$ of HII regions tends to be positively correlated with the distance from the Galactic Center ($R_{GC}$) (Churchwell et al. 1978; Downes et al. 1980; Wink et al. 1983; Shaver et al. 1983; Paladini et al. 2004). This is because that inner Galaxy tends to have higher metallicities, resulting in greater cooling rates and lower $T_e$ of HII region (Shaver et al. 1983, Quireza et al. 2006, Balser et al. 2011). However, most of these studies were conducted with single-dishes, which cannot resolve distant HII regions. These relations need to be tested with high resolution data.

⋆ E-mail: zhufy@zhejianglab.com
† E-mail: liutie@shao.ac.cn

Observations of hydrogen recombination lines at radio and (sub)millimetre (submm/mm) frequencies are routinely used by observers to infer densities, temperatures and velocity structures inside HII regions (Gordon & Sorochenko 2002). For young Ultra-Compact (UC) HII regions that are still embedded in their natal gas clumps and are not visible at optical and infrared wavelengths, submm/mm RRLs are among the best tracers for probing their physical properties because they are much less affected by the pressure broadening, and brighter than the centimeter RRLs with larger principal quantum numbers (Keto et al. 2008).

Previous statistical observations of submm/mm RRLs toward UC HII regions were mostly conducted with single dishes (Churchwell et al. 2010; Kim et al. 2017, 2018; Nguyen-Luong et al. 2017). These observations, however, cannot resolve the structures as well as gas properties of UC HII regions. In this work, we present the first systematic investigation of the properties of a large sample of resolved UC HII regions with high resolution (∼2.5″) RRLs H40$\alpha$ and He40$\alpha$ as well as the 3mm continuum emission that was obtained from the ATOMS (ALMA Three-millimeter Observations of Massive Star forming regions) survey. In particular, we focus on the measurements of $n_{He^+}/n_{H^+}$ number density ratios and electron temperatures of these UC HII regions.

In section 2, we describe the observations. Section 3 presents the results, focusing on statistics of parameters such as electron density ($n_e$), emission measure (EM), $T_e$, $n_{He^+}/n_{H^+}$ ratio and line width. A discussion of the results is given in Section 4. We summarize our main conclusions in Section 5.

## 2 OBSERVATIONS

### 2.1 The sample

We made the ALMA observations for 146 active Galactic star forming regions through the ATOMS (ALMA Three-millimeter Observations of Massive Star forming regions) survey. These 146 sources were selected from the CS J = 2-1 survey of Bronfman et al. (1996), a complete and homogeneous molecular line survey of UC HII region candidates in the Galactic plane. The sample of 146 targets is complete for proto-clusters with bright CS J = 2-1 emission ($T_b$ >2 K), indicative of rather dense gas. The overall properties of these 146 targets are shown in Table A of Liu et al. (2020a). Majority (139) of the targets are located in the first and fourth Galactic Quadrants of the inner Galactic Plane. The distances of the sample clouds range from 0.4 kpc to 13.0 kpc with a mean value of 4.5 kpc. The sample includes 27 distant (d>7 kpc) sources that are either close to the Galactic center or mini-starbursts, representing extreme environments for star formation. The properties of these sources have been described in detail in Liu et al. (2016, 2020a,b) and Liu et al. (2021).

### 2.2 ALMA observations

The ATOMS data consist of 146 sources observed in both ALMA-12m and ACA-7m arrays. The spectral setup consists of 8 spectral windows (spw), where spws 1-6 are narrow bands centered on particular lines, and spws 7-8 are broad band (1.875 GHz), low spectral resolution for continuum and lines. Table 2 of Liu et al. 2020a summerizes the details of receiver setups.

Fully calibrated visibility data are generated by running the ALMA Pipeline, which applies calibration tables obtained from the QA2 to the raw data. Images are made from the combined 12m and ACA visibility data. Continuum images are constructed from line-free channels in spw 7 and 8 centered at ∼ 99.4 GHz (or 3 mm). Line images are made for each spw with native spectral resolution. More details of data reduction are described in Liu et al. (2022), Zhou et al. (2021) and Wang Ke et al. (in prep). Uncertainties in flux calibration are estimated to be ∼ 10%. All images are primary-beam corrected. The typical rms noise (1 sigma) for the continuum images is ∼0.2 mJy for a synthesized beam of ∼ 2″. In this work, we also use RRLs H40$\alpha$ and He40$\alpha$ line data with spectral resolution of 1.5 km s$^{-1}$. The typical beam size and channel rms noise level for RRLs are ∼ 2.5 ″ and 0.02 Jy beam$^{-1}$, respectively. The rest frequencies of H40$\alpha$ and He40$\alpha$ line are 99.022952 and 99.063305 GHz, respectively.

## 3 RESULT

### 3.1 Resolved UC HII regions identified with RRL H40$\alpha$

The raw data of RRL and 3mm observations are cubes and continuum images, respectively. We integrate the RRL cube along the velocity axis to get intensity maps. Then we extracted compact objects (UC HII regions) from the integrated intensity maps of the H40$\alpha$ and He40$\alpha$. Figure 1 shows the integrated intensity maps for two exemplar sources I15567-5236 and I09002-4732. In each source, the left and middle panels show the integrated intensity maps of H40$\alpha$ and He40$\alpha$, respectively. The 3 mm continuum emission is shown as contours in the left panel. The 3 mm continuum emission coincides with the RRLs emission very well, indicating that the 3 mm continuum is dominated by free-free emission, as also suggested in Keto et al. (2008).

The compact objects in H40$\alpha$ and He40$\alpha$ emission can be easily identified by eyes. In total, we detected 94 and 44 compact sources (UC HII regions) in emission from H40$\alpha$ and He40$\alpha$, respectively. The number of the UC HII regions agrees with that by Liu et al. (2021). Table C1 summarized the parameters for H40$\alpha$, He40$\alpha$ and 3mm emission. The majority (∼90%) of UC HII regions are resolved. In further analysis, we are neglecting unresolved HII regions (with sizes smaller than one beam), which are labeled ∗ after the IRAS name in Table 1 and Table C1 .

The majority (∼86%) of targeted sources contain only one UC HII region in RRLs line emission as illustrated for I09002-4732 in the bottom panel of Figure 1. For the case of 'mutiple sources' which are very close to each other and hard to divide, like I15567-5236 in the top panel of Figure 1, we treated them as one single UC HII region. However, we identified multiple UC HII regions in 10 sources from H40$\alpha$ maps. We derive parameters for individual UC HII regions and listed them in Table 1. The effective radius ($R_{eff}$) of each UC HII region is defined as $R_{eff} = \sqrt{S/\pi}$, where S is the area of the UC HII region within the 5$\sigma$ contours. $R_{eff}$ are listed in the third column of Table 1. The median $R_{eff}$ is ∼0.1 pc. For resolved objects, the uncertainties of $R_{eff}$ are mainly determined by distances. The uncertainties of $R_{eff}$ caused by measured area (S) are negligible.

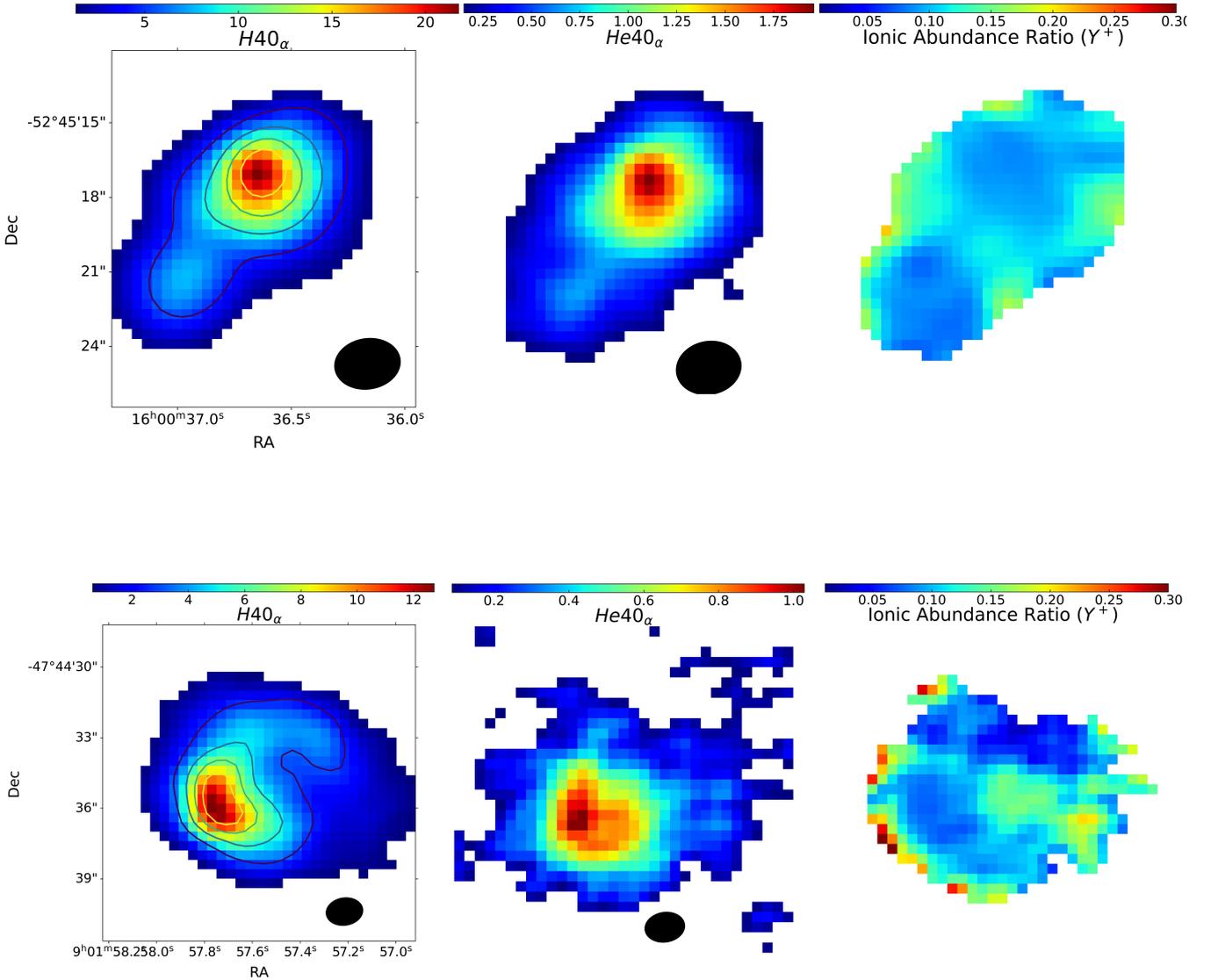

**Figure 1.** The source of I15567-5236 (top) and I09002-4732(bottom). The value of different colors show in the colorbar which in the top of each panel. For each source, the left and middle panels are H40$\alpha$ and He40$\alpha$ intensity map over 5$\sigma$, respectively. The 3 mm continuum emission is shown in contours in intensity map of H40$\alpha$. Contours are from 20% to 80% in step of 20% of peak values. The right panel shows the ionic abundance ration of the source. The black ellipses represent the synthesized beam sizes.

However, the uncertainties of distances are not well known. Therefore, we did not consider the uncertainties of $R_{\text{eff}}$. We note that this should affect the error analysis of some derived parameteres (such as emission measure and electron density) in following analysis, which can be better dealt with from more accurate distance measurements in future.

Figure 2 presents the correlation between distance and $R_{\text{eff}}$ of the sample. There is a noticeable increasing trend between the distance and $R_{\text{eff}}$. The correlation coefficient is 0.45 with p-value $\ll$ 0.001. The increasing trend indicates that in distant sources only large ($R_{\text{eff}} > 0.1$ pc) UC H II regions were resolved in ATOMS observations.

### 3.2 Properties of ionized Gas within UC H II regions

#### 3.2.1 Emission measure and electron density derived from H40$\alpha$ emission

To derive emission measure (EM) and electron density ($n_e$) from an RRL, we assume that the total rate of recombinations can be computed from $R_{\text{rec}} = n_e n_p V \alpha_B$, where $n_p$ is the proton density. $V$ is the volume and $\alpha_B$ is the total recombination coefficient excluding captures to the n=1 levels (i.e., case B). The ionization of helium and stimulation of RRLs are also assumed to be negligible. Following Zhang et al. (2022), we define the line luminosity (in CGS units) as:

$$L_\nu(\text{H}40\alpha) = n_e n_p V \epsilon, \quad (1)$$

**Table 1.** Gas clumps identified in H40α.

| IRAS | ID | $R_{\text{eff}}$ (pc) | $R_{\text{GC}}$ (kpc) | vlsr (km/s) | EM ($10^6$ cm$^{-6}$ pc) RRL | $n_e$ ($10^4$ cm$^{-3}$) RRL | EM ($10^6$ cm$^{-6}$ pc) 3mm | $n_e$ ($10^4$ cm$^{-3}$) 3mm | Y$^+$ | Temperature (K) |
|---|---|---|---|---|---|---|---|---|---|---|
| I09002-4732 | 1 | 0.03 | 8.4 | 3.1 | 83.01 ± 8.31 | 3.67 ± 0.18 | 108.77 ± 9.75 | 4.19 ± 0.19 | 0.11 ± 0.01 | 7570 ± 790 |
| I11298-6155 | 1 | 0.25 | 10.1 | 32.9 | 3.04 ± 0.30 | 0.25 ± 0.01 | 4.94 ± 0.41 | 0.32 ± 0.01 | 0.00 ± 0.00 | 7270 ± 690 |
| I12320-6122 | 1 | 0.06 | 7.2 | -42.5 | 27.93 ± 2.80 | 1.54 ± 0.08 | 34.67 ± 2.52 | 1.71 ± 0.06 | 0.07 ± 0.01 | 7040 ± 600 |
| I12326-6245 | 1 | 0.06 | 7.2 | -39.6 | 150.57 ± 15.15 | 3.56 ± 0.18 | 285.82 ± 21.72 | 4.92 ± 0.19 | 0.09 ± 0.01 | 9000 ± 850 |
| I12383-6128 | 1 | 0.07 | 7.2 | -39.1 | 3.79 ± 0.38 | 0.52 ± 0.03 | 4.75 ± 0.34 | 0.58 ± 0.02 | 0.00 ± 0.00 | 6560 ± 560 |
| I12572-6316 | 1 | 0.16 | 9.8 | 30.9 | 2.62 ± 0.26 | 0.29 ± 0.01 | 6.36 ± 0.52 | 0.45 ± 0.02 | 0.00 ± 0.00 | 11060 ± 940 |
| I13080-6229 | 1 | 0.25 | 6.9 | -35.6 | 8.27 ± 0.83 | 0.41 ± 0.02 | 11.79 ± 1.31 | 0.49 ± 0.03 | 0.14 ± 0.02 | 7480 ± 860 |
| I13111-6228 | 1 | 0.07 | 6.9 | -38.8 | 2.10 ± 0.21 | 0.38 ± 0.02 | 3.10 ± 0.41 | 0.46 ± 0.03 | 0.00 ± 0.00 | 8420 ± 720 |
| I13291-6229 | 1 | 0.08 | 7.0 | -36.5 | 3.10 ± 0.31 | 0.45 ± 0.02 | 2.85 ± 0.22 | 0.43 ± 0.02 | 0.00 ± 0.00 | 5000 ± 420 |
| I13291-6229 | 2 | 0.08 | 7.0 | -36.5 | 1.91 ± 0.19 | 0.34 ± 0.02 | 2.22 ± 0.17 | 0.37 ± 0.01 | 0.00 ± 0.00 | 5810 ± 610 |
| I13291-6249 | 1 | 0.26 | 7.1 | -34.7 | 6.95 ± 0.70 | 0.37 ± 0.02 | 10.57 ± 0.91 | 0.45 ± 0.02 | 0.14 ± 0.02 | 7300 ± 910 |
| I13471-6120* | 1 | 0.07 | 6.4 | -56.7 | 79.29 ± 7.95 | 2.33 ± 0.12 | 85.02 ± 7.37 | 2.41 ± 0.10 | 0.09 ± 0.01 | 6500 ± 620 |
| I14013-6105 | 1 | 0.11 | 6.4 | -48.1 | 24.59 ± 2.46 | 1.04 ± 0.05 | 33.70 ± 3.58 | 1.22 ± 0.06 | 0.12 ± 0.01 | 7290 ± 840 |
| I14050-6056 | 1 | 0.15 | 6.6 | -47.1 | 7.20 ± 0.72 | 0.48 ± 0.02 | 8.87 ± 1.05 | 0.53 ± 0.03 | 0.16 ± 0.02 | 6310 ± 850 |
| I14382-6017 | 1 | 0.57 | 6.0 | -60.7 | 2.52 ± 0.25 | 0.15 ± 0.01 | 2.99 ± 0.33 | 0.16 ± 0.01 | 0.00 ± 0.00 | 6470 ± 620 |
| I14453-5912 | 1 | 0.19 | 6.6 | -40.2 | 1.31 ± 0.13 | 0.19 ± 0.01 | 1.41 ± 0.15 | 0.19 ± 0.01 | 0.00 ± 0.00 | 5330 ± 510 |
| I15254-5621 | 1 | 0.06 | 5.7 | -67.3 | 98.39 ± 9.87 | 2.88 ± 0.14 | 111.57 ± 8.74 | 3.05 ± 0.12 | 0.09 ± 0.01 | 7140 ± 680 |
| I15290-5546 | 1 | 0.11 | 4.9 | -87.5 | 27.35 ± 2.74 | 1.09 ± 0.05 | 43.16 ± 3.45 | 1.38 ± 0.06 | 0.10 ± 0.01 | 7570 ± 790 |
| I15290-5546 | 2 | 0.21 | 4.9 | -87.5 | 23.78 ± 2.38 | 0.74 ± 0.04 | 27.35 ± 2.83 | 0.80 ± 0.04 | 0.11 ± 0.01 | 6920 ± 730 |
| I15384-5348 | 1 | 0.10 | 6.9 | -41.0 | 7.04 ± 0.70 | 0.60 ± 0.03 | 9.17 ± 1.09 | 0.69 ± 0.04 | 0.14 ± 0.02 | 6490 ± 880 |
| I15408-5356 | 1 | 0.14 | 6.9 | -39.7 | 6.10 ± 0.61 | 0.47 ± 0.02 | 7.16 ± 1.38 | 0.51 ± 0.05 | 0.18 ± 0.02 | 7580 ± 1020 |
| I15411-5352 | 1 | 0.10 | 6.9 | -41.5 | 15.51 ± 1.55 | 0.90 ± 0.05 | 18.84 ± 2.11 | 1.00 ± 0.06 | 0.12 ± 0.01 | 6640 ± 760 |
| I15439-5449 | 1 | 0.06 | 5.9 | -54.6 | 13.71 ± 1.37 | 1.06 ± 0.05 | 18.66 ± 1.39 | 1.24 ± 0.05 | 0.00 ± 0.00 | 7040 ± 600 |
| I15502-5302 | 1 | 0.10 | 4.6 | -91.4 | 238.75 ± 23.93 | 3.55 ± 0.18 | 285.80 ± 23.75 | 3.88 ± 0.16 | 0.11 ± 0.01 | 8640 ± 940 |
| I15520-5234 | 1 | 0.07 | 6.2 | -41.3 | 52.62 ± 5.27 | 1.91 ± 0.10 | 49.02 ± 3.67 | 1.85 ± 0.07 | 0.00 ± 0.00 | 5840 ± 500 |
| I15567-5236 | 1 | 0.15 | 4.4 | -107.1 | 65.14 ± 6.52 | 1.49 ± 0.07 | 77.70 ± 6.76 | 1.63 ± 0.07 | 0.11 ± 0.01 | 6650 ± 700 |
| I15570-5227 | 1 | 0.31 | 4.4 | -101.5 | 3.37 ± 0.34 | 0.23 ± 0.01 | 4.22 ± 0.35 | 0.26 ± 0.01 | 0.00 ± 0.00 | 6430 ± 550 |
| I16037-5223* | 1 | 0.12 | 4.9 | -80.0 | 43.34 ± 4.36 | 1.35 ± 0.07 | 51.39 ± 4.25 | 1.47 ± 0.06 | 0.11 ± 0.01 | 6430 ± 670 |
| I16037-5223* | 2 | 0.14 | 4.9 | -80.0 | 22.70 ± 2.28 | 0.91 ± 0.05 | 27.44 ± 2.34 | 1.00 ± 0.04 | 0.09 ± 0.02 | 6590 ± 820 |
| I16037-5223* | 3 | 0.12 | 4.9 | -80.0 | 12.40 ± 1.25 | 0.70 ± 0.04 | 12.84 ± 1.19 | 0.72 ± 0.03 | 0.11 ± 0.02 | 5860 ± 730 |
| I16060-5146 | 1 | 0.09 | 4.5 | -91.6 | 403.90 ± 40.48 | 4.82 ± 0.24 | 539.91 ± 43.10 | 5.57 ± 0.22 | 0.10 ± 0.01 | 10060 ± 970 |
| I16065-5158 | 1 | 0.11 | 5.2 | -63.3 | 10.39 ± 1.04 | 0.70 ± 0.04 | 23.15 ± 1.74 | 1.05 ± 0.04 | 0.00 ± 0.00 | 9250 ± 880 |
| I16065-5158 | 2 | 0.07 | 5.2 | -63.3 | 7.62 ± 0.77 | 0.73 ± 0.04 | 11.77 ± 0.70 | 0.90 ± 0.03 | 0.00 ± 0.00 | 5990 ± 630 |
| I16071-5142 | 1 | 0.10 | 4.5 | -87.0 | 7.63 ± 0.77 | 0.60 ± 0.03 | 16.54 ± 1.23 | 0.89 ± 0.03 | 0.00 ± 0.00 | 8930 ± 760 |
| I16132-5039 | 1 | 0.09 | 5.8 | -47.5 | 1.83 ± 0.18 | 0.32 ± 0.02 | 1.53 ± 0.07 | 0.30 ± 0.01 | 0.00 ± 0.00 | 4230 ± 400 |
| I16158-5055 | 1 | 0.10 | 5.4 | -49.2 | 3.44 ± 0.34 | 0.40 ± 0.02 | 4.31 ± 0.69 | 0.45 ± 0.04 | 0.00 ± 0.00 | 6500 ± 620 |
| I16164-5046 | 1 | 0.06 | 5.4 | -57.3 | 306.24 ± 30.72 | 5.06 ± 0.25 | 349.31 ± 27.77 | 5.40 ± 0.21 | 0.08 ± 0.01 | 6970 ± 660 |
| I16172-5028 | 1 | 0.06 | 5.4 | -51.9 | 201.51 ± 20.19 | 4.06 ± 0.20 | 245.53 ± 20.05 | 4.49 ± 0.18 | 0.12 ± 0.01 | 8400 ± 900 |
| I16177-5018 | 1 | 0.05 | 5.4 | -50.2 | 43.19 ± 4.35 | 2.11 ± 0.11 | 63.76 ± 3.59 | 2.58 ± 0.07 | 0.11 ± 0.02 | 6610 ± 830 |
| I16177-5018 | 2 | 0.08 | 5.4 | -50.2 | 39.48 ± 3.95 | 1.54 ± 0.08 | 58.31 ± 2.59 | 1.87 ± 0.04 | 0.10 ± 0.02 | 6910 ± 860 |
| I16177-5018 | 3 | 0.03 | 5.4 | -50.2 | 43.76 ± 4.41 | 2.76 ± 0.14 | 66.06 ± 4.10 | 3.37 ± 0.10 | 0.11 ± 0.02 | 6660 ± 900 |
| I16177-5018 | 4 | 0.08 | 5.4 | -50.2 | 15.15 ± 1.52 | 0.95 ± 0.05 | 21.75 ± 1.98 | 1.14 ± 0.05 | 0.15 ± 0.02 | 6490 ± 880 |
| I16177-5018 | 5 | 0.12 | 5.4 | -50.2 | 19.52 ± 1.95 | 0.92 ± 0.05 | 36.00 ± 3.64 | 1.25 ± 0.06 | 0.10 ± 0.02 | 6790 ± 780 |
| I16297-4757 | 1 | 0.20 | 4.2 | -79.6 | 3.26 ± 0.33 | 0.28 ± 0.01 | 3.55 ± 0.35 | 0.29 ± 0.01 | 0.00 ± 0.00 | 6010 ± 570 |
| I16304-4710 | 1 | 0.27 | 4.9 | -62.8 | 5.95 ± 0.60 | 0.33 ± 0.02 | 6.72 ± 0.53 | 0.35 ± 0.01 | 0.00 ± 0.00 | 6280 ± 600 |
| I16313-4729* | 1 | 0.06 | 4.4 | -73.7 | 8.70 ± 0.88 | 0.88 ± 0.04 | 13.21 ± 0.99 | 1.09 ± 0.04 | 0.00 ± 0.00 | 7660 ± 730 |
| I16318-4724 | 1 | 0.19 | 3.3 | -119.8 | 3.84 ± 0.39 | 0.32 ± 0.02 | 4.18 ± 0.43 | 0.33 ± 0.02 | 0.00 ± 0.00 | 5960 ± 620 |
| I16330-4725 | 1 | 0.18 | 4.6 | -75.1 | 22.73 ± 2.28 | 0.80 ± 0.04 | 32.39 ± 2.41 | 0.95 ± 0.04 | 0.00 ± 0.00 | 6820 ± 650 |
| I16330-4725 | 2 | 0.21 | 4.6 | -75.1 | 11.65 ± 1.17 | 0.53 ± 0.03 | 18.56 ± 1.53 | 0.66 ± 0.03 | 0.00 ± 0.00 | 7120 ± 750 |
| I16348-4654* | 1 | 0.15 | 5.4 | -46.5 | 40.44 ± 4.06 | 1.18 ± 0.06 | 63.09 ± 4.68 | 1.47 ± 0.05 | 0.00 ± 0.00 | 8220 ± 780 |
| I16351-4722* | 1 | 0.03 | 5.7 | -41.4 | 24.24 ± 2.44 | 2.01 ± 0.10 | 49.37 ± 3.28 | 2.87 ± 0.10 | 0.00 ± 0.00 | 8070 ± 770 |
| I16385-4619 | 1 | 0.13 | 3.1 | -117.0 | 10.32 ± 1.03 | 0.62 ± 0.03 | 11.15 ± 0.81 | 0.65 ± 0.02 | 0.00 ± 0.00 | 5740 ± 540 |
| I16445-4459 | 1 | 0.14 | 2.8 | -121.3 | 6.82 ± 0.68 | 0.50 ± 0.02 | 8.51 ± 0.60 | 0.56 ± 0.02 | 0.00 ± 0.00 | 6210 ± 590 |
| I16458-4512* | 1 | 0.05 | 5.1 | -50.4 | 20.52 ± 2.06 | 1.47 ± 0.07 | 21.07 ± 1.71 | 1.50 ± 0.06 | 0.00 ± 0.00 | 6070 ± 580 |
| I16506-4512 | 1 | 0.21 | 6.1 | -26.2 | 3.54 ± 0.35 | 0.29 ± 0.01 | 4.06 ± 0.53 | 0.31 ± 0.02 | 0.00 ± 0.00 | 6670 ± 630 |

**Table 1** – *continued*

| IRAS | ID | $R_{\rm eff}$ (pc) | $R_{\rm GC}$ (kpc) | vlsr (km/s) | EM ($10^6$ cm$^{-6}$ pc) RRL | $n_e$ ($10^4$ cm$^{-3}$) RRL | EM ($10^6$ cm$^{-6}$ pc) 3mm | $n_e$ ($10^4$ cm$^{-3}$) 3mm | $Y^+$ | Temperature (K) |
|---|---|---|---|---|---|---|---|---|---|---|
| I17006-4215 | 1 | 0.08 | 6.3 | -23.2 | 14.15 ± 1.42 | 0.94 ± 0.05 | 13.66 ± 1.38 | 0.92 ± 0.05 | 0.10 ± 0.01 | 6120 ± 640 |
| I17016-4124* | 1 | 0.02 | 7.0 | -27.1 | 43.56 ± 4.37 | 3.58 ± 0.18 | 74.99 ± 6.22 | 4.70 ± 0.19 | 0.00 ± 0.00 | 8010 ± 680 |
| I17136-3617 | 1 | 0.04 | 7.0 | -10.6 | 41.83 ± 4.19 | 2.34 ± 0.12 | 51.71 ± 3.94 | 2.61 ± 0.10 | 0.08 ± 0.01 | 6880 ± 650 |
| I17143-3700 | 1 | 0.22 | 4.7 | -31.1 | 11.36 ± 1.14 | 0.51 ± 0.03 | 13.10 ± 1.13 | 0.55 ± 0.02 | 0.13 ± 0.01 | 6210 ± 710 |
| I17160-3707 | 1 | 0.62 | 2.7 | -69.5 | 6.06 ± 0.61 | 0.22 ± 0.01 | 6.12 ± 0.72 | 0.22 ± 0.01 | 0.00 ± 0.00 | 6180 ± 590 |
| I17175-3544 | 1 | 0.04 | 7.0 | -5.7 | 79.67 ± 7.97 | 3.26 ± 0.16 | 83.59 ± 7.04 | 3.32 ± 0.14 | 0.00 ± 0.00 | 6780 ± 580 |
| I17204-3636 | 1 | 0.06 | 5.1 | -18.2 | 6.51 ± 0.65 | 0.76 ± 0.04 | 9.42 ± 0.79 | 0.92 ± 0.04 | 0.00 ± 0.00 | 6410 ± 670 |
| I17220-3609 | 1 | 0.13 | 1.3 | -93.7 | 63.37 ± 6.35 | 1.57 ± 0.08 | 76.06 ± 6.50 | 1.72 ± 0.07 | 0.11 ± 0.01 | 6720 ± 700 |
| I17233-3606 | 1 | 0.03 | 7.0 | -2.7 | 14.13 ± 1.42 | 1.63 ± 0.08 | 12.85 ± 0.91 | 1.57 ± 0.06 | 0.00 ± 0.00 | 5060 ± 430 |
| I17244-3536 | 1 | 0.03 | 7.0 | -10.2 | 5.54 ± 0.55 | 0.93 ± 0.05 | 5.87 ± 0.53 | 0.96 ± 0.04 | 0.00 ± 0.00 | 6610 ± 560 |
| I17258-3637 | 1 | 0.13 | 5.8 | -11.9 | 71.44 ± 7.15 | 1.63 ± 0.08 | 83.80 ± 7.84 | 1.77 ± 0.08 | 0.12 ± 0.01 | 6720 ± 700 |
| I17271-3439 | 1 | 0.06 | 5.3 | -18.2 | 28.63 ± 2.87 | 1.57 ± 0.08 | 32.06 ± 2.67 | 1.66 ± 0.07 | 0.00 ± 0.00 | 6380 ± 480 |
| I17441-2822* | 1 | 0.08 | 0.2 | 50.8 | 759.17 ± 76.67 | 6.95 ± 0.35 | 1613.45 ± 118.76 | 10.11 ± 0.37 | 0.07 ± 0.01 | 12110 ± 1330 |
| I17441-2822* | 2 | 0.10 | 0.2 | 50.8 | 333.11 ± 33.60 | 4.03 ± 0.20 | 439.87 ± 31.63 | 4.62 ± 0.17 | 0.07 ± 0.01 | 6850 ± 790 |
| I17455-2800 | 1 | 0.55 | 1.7 | -15.6 | 4.84 ± 0.48 | 0.21 ± 0.01 | 5.00 ± 0.69 | 0.21 ± 0.01 | 0.21 ± 0.03 | 5610 ± 870 |
| I17545-2357* | 1 | 0.04 | 5.4 | 7.9 | 15.64 ± 1.57 | 1.47 ± 0.07 | 16.69 ± 1.15 | 1.52 ± 0.05 | 0.00 ± 0.00 | 6220 ± 530 |
| I17599-2148 | 1 | 0.05 | 5.4 | 18.6 | 6.26 ± 0.63 | 0.76 ± 0.04 | 10.53 ± 0.94 | 0.99 ± 0.04 | 0.00 ± 0.00 | 7450 ± 780 |
| I17599-2148 | 2 | 0.06 | 5.4 | 18.6 | 10.55 ± 1.06 | 0.93 ± 0.05 | 17.26 ± 1.47 | 1.19 ± 0.05 | 0.00 ± 0.00 | 7210 ± 760 |
| I17599-2148 | 3 | 0.06 | 5.4 | 18.6 | 9.85 ± 0.99 | 0.94 ± 0.05 | 15.09 ± 1.40 | 1.16 ± 0.05 | 0.00 ± 0.00 | 6660 ± 700 |
| I18032-2032* | 1 | 0.05 | 3.4 | 4.3 | 16.17 ± 1.63 | 1.23 ± 0.06 | 24.12 ± 1.88 | 1.51 ± 0.06 | 0.00 ± 0.00 | 6640 ± 560 |
| I18110-1854 | 1 | 0.10 | 5.1 | 37.0 | 20.15 ± 2.02 | 0.99 ± 0.05 | 20.36 ± 2.02 | 1.00 ± 0.05 | 0.10 ± 0.01 | 6350 ± 670 |
| I18116-1646 | 1 | 0.28 | 4.6 | 48.5 | 5.32 ± 0.53 | 0.31 ± 0.02 | 6.14 ± 0.65 | 0.33 ± 0.02 | 0.00 ± 0.00 | 6360 ± 540 |
| I18139-1842 | 1 | 0.06 | 5.4 | 39.8 | 4.00 ± 0.40 | 0.56 ± 0.03 | 4.83 ± 0.34 | 0.61 ± 0.02 | 0.00 ± 0.00 | 5900 ± 500 |
| I18228-1312 | 1 | 0.07 | 5.4 | 32.3 | 9.31 ± 0.93 | 0.83 ± 0.04 | 11.98 ± 0.85 | 0.94 ± 0.03 | 0.00 ± 0.00 | 6150 ± 580 |
| I18311-0809 | 1 | 0.16 | 3.7 | 113.0 | 5.73 ± 0.57 | 0.43 ± 0.02 | 6.67 ± 0.62 | 0.46 ± 0.02 | 0.00 ± 0.00 | 6280 ± 600 |
| I18314-0720 | 1 | 0.52 | 3.9 | 101.5 | 1.96 ± 0.20 | 0.14 ± 0.01 | 1.19 ± 0.20 | 0.11 ± 0.01 | 0.00 ± 0.00 | 5400 ± 460 |
| I18317-0757 | 1 | 0.25 | 4.4 | 80.7 | 7.70 ± 0.77 | 0.39 ± 0.02 | 7.24 ± 0.71 | 0.38 ± 0.02 | 0.00 ± 0.00 | 5600 ± 480 |
| I18434-0242 | 1 | 0.25 | 4.7 | 97.2 | 20.78 ± 2.08 | 0.65 ± 0.03 | 20.69 ± 2.18 | 0.65 ± 0.03 | 0.13 ± 0.01 | 6250 ± 720 |
| I18479-0005 | 1 | 0.16 | 7.5 | 14.6 | 145.15 ± 14.58 | 2.13 ± 0.11 | 215.65 ± 18.35 | 2.59 ± 0.11 | 0.09 ± 0.01 | 7720 ± 890 |
| I18479-0005 | 2 | 0.13 | 7.5 | 14.6 | 70.49 ± 7.12 | 1.65 ± 0.08 | 117.27 ± 6.97 | 2.12 ± 0.06 | 0.11 ± 0.01 | 7530 ± 940 |
| I18479-0005 | 3 | 0.19 | 7.5 | 14.6 | 39.86 ± 4.00 | 1.02 ± 0.05 | 86.43 ± 8.12 | 1.51 ± 0.07 | 0.12 ± 0.02 | 7580 ± 1020 |
| I18507p0110 | 1 | 0.02 | 7.1 | 57.2 | 429.07 ± 42.99 | 9.55 ± 0.48 | 597.33 ± 50.43 | 11.16 ± 0.47 | 0.00 ± 0.00 | 9270 ± 1080 |
| I18530p0215 | 1 | 0.12 | 5.3 | 74.1 | 7.33 ± 0.73 | 0.56 ± 0.03 | 7.64 ± 0.68 | 0.58 ± 0.03 | 0.00 ± 0.00 | 5560 ± 580 |
| I19078p0901 | 1 | 0.17 | 7.6 | 2.9 | 252.93 ± 25.40 | 2.71 ± 0.14 | 440.42 ± 37.65 | 3.58 ± 0.15 | 0.09 ± 0.01 | 10590 ± 1200 |
| I19078p0901 | 2 | 0.10 | 7.6 | 2.9 | 236.05 ± 23.87 | 3.41 ± 0.17 | 473.53 ± 35.59 | 4.84 ± 0.18 | 0.08 ± 0.01 | 9820 ± 1270 |
| I19078p0901 | 3 | 0.09 | 7.6 | 2.9 | 188.82 ± 19.13 | 3.32 ± 0.17 | 343.64 ± 28.59 | 4.47 ± 0.19 | 0.08 ± 0.01 | 9340 ± 990 |
| I19078p0901 | 4 | 0.14 | 7.6 | 2.9 | 65.85 ± 6.61 | 1.56 ± 0.08 | 109.36 ± 9.88 | 2.01 ± 0.09 | 0.08 ± 0.01 | 7880 ± 750 |
| I19095p0930 | 1 | 0.07 | 5.8 | 43.7 | 17.60 ± 1.77 | 1.10 ± 0.06 | 36.83 ± 2.43 | 1.59 ± 0.05 | 0.00 ± 0.00 | 9930 ± 840 |
| I19097p0847 | 1 | 0.28 | 6.2 | 58.0 | 2.64 ± 0.26 | 0.22 ± 0.01 | 2.50 ± 0.26 | 0.21 ± 0.01 | 0.00 ± 0.00 | 5280 ± 500 |

The ∗ symbol after the IRAS name indicates that the sources are unresolved. The uncertainties of EM, $n_e$, $Y^+$ and temperature are calculated by the law of propagation of uncertainties from measured line intensities listed in Table C1.

where $\epsilon$ is the efficiency for producing H40$\alpha$ photons per recombination. Following Zhu et al. (2019), $\epsilon = 1.99 \times 10^{-32}$ at $T_e = 10^4$ K and $n_e = 10^4$ cm$^{-3}$, which are the typical temperature and density of H II regions. The variations of $\epsilon$ is about 10 per cent for densities as low as 100 cm$^3$ and temperatures as low as 5000 K (Zhang et al. 2022). For the sources with detections of H40$\alpha$, we can derive $n_e$ in unit of cm$^{-3}$:

$$n_e = 3.582 \left[\frac{L(\text{H}40\alpha)}{\text{Jy km/s kpc}^2}\right]^{0.5} \left[\frac{R_{\rm eff}}{\text{pc}}\right]^{-1.5} \quad (2)$$

where $L_\nu(\text{H}40\alpha) = L(\text{H}40\alpha)\nu/c$ is the line luminosity in observational units (Jy km s$^{-1}$ kpc$^2$). We assume an H II region in our sample to be a homogeneous medium with a thickness of $2R_{\rm eff}$, then the relation between $n_e$ and EM is:

$$EM = 2R_{\rm eff} n_e n_p \quad (3)$$

$$EM = 25.66 \frac{L(\text{H}40\alpha)}{\text{Jy km/s kpc}^2} \left[\frac{R_{\rm eff}}{\text{pc}}\right]^{-2} \quad (4)$$

where EM in unit cm$^{-6}$ pc.

The median, mean and standard deviation of EM are $1.5 \times 10^7$, $5.7 \times 10^7$ and $1.1 \times 10^8$ cm$^{-6}$ pc, respectively. The median, mean and standard deviation of $n_e$ are $9.2 \times 10^3$, $1.4 \times 10^4$ and $1.5 \times 10^4$ cm$^{-3}$, respectively. The EM and $n_e$ are listed in the sixth and seventh column of Table 1.

### 3.2.2 Ionic Abundance Ratio

Here we present measurements of the ionized helium-to-hydrogen ratio ($Y^+ = n_{\text{He}^+}/n_{\text{H}^+}$) toward UC H II regions. We derive $Y^+$ within

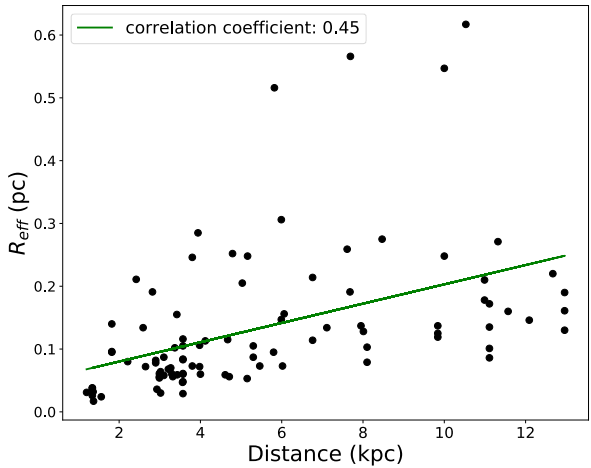

**Figure 2.** The correlation between distance and $R_{\text{eff}}$. The green line shows the linear fitting.

UC HII regions pixel-by-pixel using (Luisi et al. 2016; Roshi et al. 2017):

$$Y^+ = \frac{\int I_{\text{He}^+} d\nu}{\int I_{\text{H}^+} d\nu} \quad (5)$$

where $\int I_{\text{He}^+} d\nu$ and $\int I_{\text{H}^+} d\nu$ are the integrated intensity in Jy beam$^{-1}$ km s$^{-1}$. Here we assume both helium and hydrogen RRLs are in the LTE condition. The right panel of Figure 1 shows the $Y^+$ map of two exemplar sources. The $Y^+$ distribution within the UC HII regions is not uniform. $Y^+$ appears to higher in the boundary region than in the center. More disscussion on $Y^+$ maps are presented in Section 4.

There are 44 sources showing He40α emission in our paper. The source-averaged $Y^+$ values of these 44 UC HII regions are summarized in the tenth column of Table 1. The left panel of Figure 3 shows the distributions of source-averaged $Y^+$. The median, mean and standard deviation of source-averaged $Y^+$ are 0.11, 0.11 and 0.03, respectively.

*3.2.3 Electron temperature*

As demonstrated in Keto et al. (2008), the observed 3 mm continuum emission of UC HII regions in high resolution interferometer observations should be mostly free-free emission rather than from dust. In Figure 4, we compare the intensity maps of H40α and H$^{13}$CO$^+$ for an example source, I09002-4732. From this figure, one can see that the UC HII region where H40α is bright shows very weak or even no H$^{13}$CO$^+$ molecular line emission, indicating that the UC HII region contains negligible cold thermal dust radiation. Therefore, its 3 mm continuum emission could mainly come from free-free emission. The other UC HII regions in our sample are similar to I09002-4732 in that regard. However, we cannot rule out the possibility of the existence of some warm/hot dust emission inside UC HII regions that cannot be traced by H$^{13}$CO$^+$ line emission.

There is the another way to show if 3 mm continuum emission should be mostly free-free emission. The amount of continuum flux in excess of the flux predicted can be used to calculate the mass of dust causing the excess. This mass is given by (Pratap et al. 1992)

$$M_{\text{dust}} = 1.91 \times 10^{-2} (\frac{\lambda_{mm}}{0.2})^{\beta+3} S_\nu [exp(\frac{14.4}{\lambda_{mm} T_d})] d_{\text{kpc}}^2 M_\odot \quad (6)$$

where $S_\nu$ is the excess flux due to emission from the dust, $\beta$ is the dust emissivity (assumed = 1 Pratap et al. 1992), $T_d$ is the dust temperature, and $d_{kpc}$ is the distance. For the exemplar source, I09002-4732, its observed flux densities ($S_\nu$) is 13.3 Jy and the $T_d$ and $d_{\text{kpc}}$ are 39 K and 1.2 kpc (Liu et al. 2020a), respectively. If there are 10% 3mm continuum emission come from dust, the corresponding gas mass derived from dust emission is ∼ 250 M$_\odot$, comparable to that of its natal clump (∼251 M$_\odot$ Liu et al. 2020a). This also indicates that only a negligible fraction (less than 10%) of 3mm continuum emission from the UC HII region is from thermal dust radiation. The other UC HII regions in our sample are similar to I09002-4732 in that regard. We note that future higher frequency continuum observations can help constrain the dust emission within UC HII regions in a better way.

Assuming that the 3 mm continuum emission is mainly from free-free emission, the electron temperatures derived from the intensities of the H40α RRL emission and the 3 mm continuum emission, using the equations given in Appendix A in non-LTE conditions, are listed in the last column of Table 1. The middle panel of Figure 3 shows the electron temperature distributions of UC HII regions. The median, mean and standard deviation of electron temperature are 6662, 7026 and 1348 K, respectively. The values of the electron temperatures are typical of HII regions, but somehow on the cool side for what's expected of UC HII regions, indicating that 3 mm continuum emission is mainly from free-free emission is reasonable.

*3.2.4 Electron density derived from 3 mm radio continuum*

With electron temperature information, electron density can also be derived from 3 mm continuum emission with equations in Appendix B. The electron densities derived from 3 mm continuum are listed in the eighth column of Table 1.

The left panel in Figure 5 shows the correlation between EM (calculated by 3mm continuum emission) and $n_e$ (calculated by RRLs). The two parameters $n_e$ and EM are strongly correlated. The slope of linear fitting is 1.55 ± 0.05. The correlation coefficient is 0.94 with p-value ≪ 0.001. The right panel shows the correlation between $n_e$ and HII region diameter. The black crosses represent the data from our paper. The gray markers are the data from Garay & Lizano (1999), which show similar trends as our data. The correlation coefficient is -0.63 with p-value ≪ 0.001. Kim et al. (2017) also found a strong negative correlation between $n_e$ and HII region diameter.

We present a comparison of the $n_e$ derived independently from the 3 mm radio continuum and H40α emission in Figure 6. There is a strong correlation between these two measurements with a correlation coefficient of 0.97 and p-value ≪ 0.001. The xy-bisector fit to these data gives a slope of 0.95 ± 0.01, indicating that millimeter RRLs are good probes of electron densities.

*3.2.5 Line widths of RRL H40α*

We calculate line widths (ΔV) of H40α from the Gaussian fit of the source-averaged spectral line. The right panel in Figure 3 shows the distribution of ΔV. The mean, median and standard deviation of ΔV are 28.06, 30.41 and 8.90 km/s, respectively. The ΔV is weakly correlated with $R_{\text{eff}}$. The correlation coefficient is -0.25 with p-value ≪ 0.001. It indicates that the H40α line width decreases as the $R_{\text{eff}}$ increases. Liu et al. (2021) found similar results and proposed non-thermal (e.g., dynamical) mechanisms dominate the line broadening of H40$\alpha$ on small scales of UC H II regions.

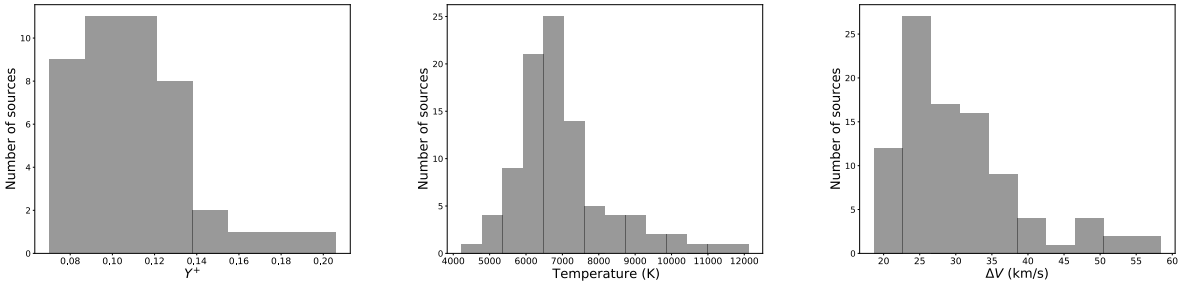

**Figure 3.** Left panel: The distribution of Y$^+$ which calculated by Equation 5. Middle panel: The distribution of Temperature which is calculated from both 3mm continuum emission and H40$\alpha$ emission. Right panel: The distribution of H40$\alpha$ emission line width. All the sources detected in the H40$\alpha$ line are used in the middel and right panels, and the sources in the left panel are detected in both H40$\alpha$ and He40$\alpha$ lines.

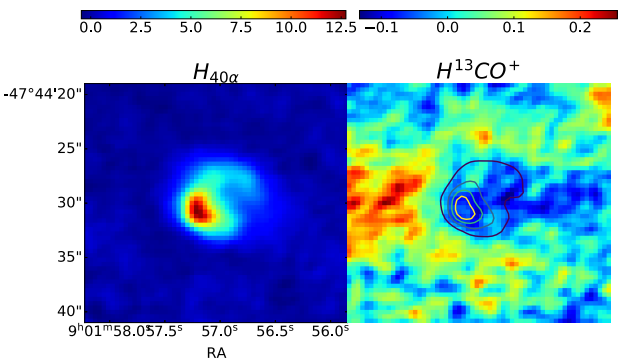

**Figure 4.** The left and right panel are the intensity of H40$\alpha$ and H$^{13}$CO$^+$, respectively. The contours in right panel is H40$\alpha$. Countours are from 20% to 80% in step of 20% of peak value.

## 4 DISCUSSION

### 4.1 Variance of ionic abundance ratio (Y$^+$) within UC H II regions

Previous studies such as Luisi et al. (2016) and Roshi et al. (2017) found Y$^+$ below 0.1 (around 0.08 and 0.06) inside H II regions. While the mean value of Y$^+$ in our example is 0.11. Luisi et al. (2016) and Roshi et al. (2017) also witnessed a decreasing trend in Y$^+$ with increasing distance from UC H II region beyond the photodissociation region. However, as shown in the right upper panel of Figure 1, we found that Y$^+$ within the resolved UC H II region I09002-4732 and I15567-5236 increase with increasing distance from the center of UC H II region. The other resolved UC H II regions in our sample also exhibit similar spatial variance in Y$^+$ similar to the example sources I09002-4732 and I15567-5236. Figure 7 shows how the Y$^+$ of our data changes over the distance from the center. We pick the 80%, 60%, 40%, 30%, 20% of the peak emission of H40$\alpha$ in H II regions as the dividing lines, and divide each source into five regions. For most sources, their boundary regions clearly show higher Y$^+$ than their inner regions. The right panel of Figure 7 shows the distribution of Y$^+$ within 80%-100% (black) and 20%-40% (red) regions, respectively. The two distributions are significantly different with a p-value of ≪ 0.001 from Kolmogorov-Smirnov test. This figure further demonstrates that outer parts of UC H II regions show higher Y$^+$.

Is the higher Y$^+$ near the edges of UC H II regions caused by low signal-to-noise (S/N) He40$\alpha$ spectra? He40$\alpha$ emission shown in Figure 1, however, is pretty strong with S/N ratios larger than 5 at the boundary of the UC H II region. To further check the data quality of He40$\alpha$ emission, we plot the H40$\alpha$ and He40$\alpha$ spectra near the boundary (marked by red box) and central region (marked by black box) in Figure 8. The spectra of RRLs H40$\alpha$ and He40$\alpha$ are clearly detected in the boundary region with high S/N. Therefore, we argue that the large Y$^+$ values near the edges of UC H II regions are not caused by noise in data.

The possible explanations for the increasing trend of Y$^+$ near the boundaries of UC H II regions could be: (1)The He in the central region has the secondary ionization which lead to a lower density of singly ionized He and weaker He40$\alpha$ line emission. However, the HeII64$\alpha$ line, which is produced by the He$^+$ ion recombined from He$^{2+}$ and free electron, is too weak to be detected in our data. (2) He40$\alpha$ emission which frequency is 99.063305 GHz and typical width around 30 km/s is contaminated by C40$\alpha$ emission, whose frequency is 99.07236032 GHz. Near the photodissociation region (PDR), CO is dissociated and ionized, and produces Carbon RRLs. The contamination of C40$\alpha$ emission in He40$\alpha$ could result in "stronger" observed He40$\alpha$ emission near PDR or the boundary of UC H II region.

The variance of Y$^+$ should be further studied through future higher sensitivity and higher resolution observations with more transitions of RRLs.

### 4.2 Line broadening mechanisms for RRLs

The line widths ($\Delta V$) of RRLs could be caused by thermal and microturbulent broadening as well as electron pressure broadening (Brocklehurst & Seaton 1972).

In Figure 9, we investigate the correlation between the line width ($\Delta V$) of H40$\alpha$ and electron temperature ($T_e$). A clear increasing trend is seen in the relation between $\Delta V$ and $T_e$. It indicates that higher electron temperature corresponds to larger line width. The correlation coefficient is 0.49 with p-value ≪ 0.001.

We calculate the thermal and electron pressure broadening follow-ing Peters et al. (2012). The left panel in Figure 10 shows the effect of thermal and pressure broadening on the spectral line widths of H40$\alpha$ at different electron density ($n_e$) and electron temperature ($T_e$). For the UC HII regions in our ATOMS sources, $n_e$ are between $10^3$ and $10^5$ cm$^{-3}$, and thus the pressure broadening has no significant effect on the spectral line widths.

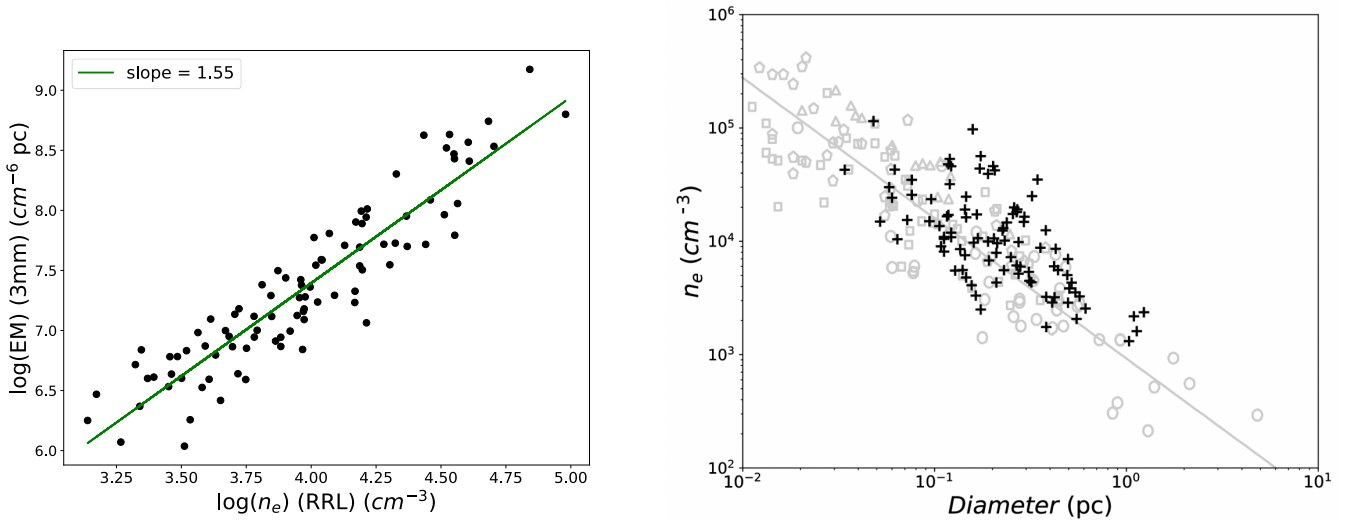

**Figure 5.** Left panel: The distribution of EM which calculated by 3mm continuum emission and $n_e$ which calculated by RRL. Right panel: The correlation between $n_e$ which calculated by RRL and HII region diameter. The black crosses represent the data from our paper. The gray markers and the line are the data from Garay & Lizano (1999) and the corresponding linear fit, respectively.

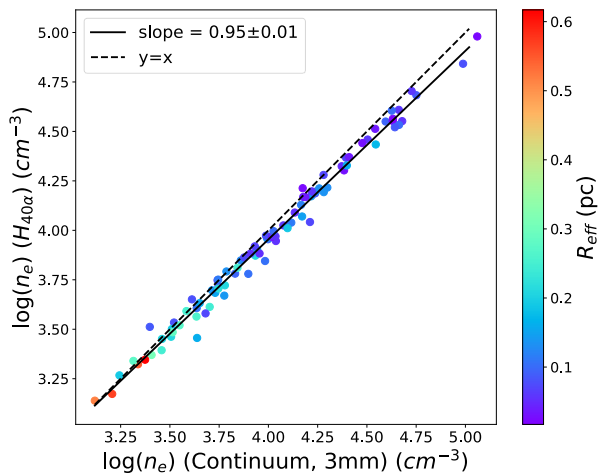

**Figure 6.** $n_e$ drived from 3 mm radio continuum emission versus $n_e$ from H40$\alpha$. The black dash line is y=x. The slope of linear fitting is 0.95 which is shown by black line.

The thermal broadening, $\Delta V_{\rm th}$, is given by $\Delta V_{\rm th} = (8ln2k_B T_e/m_H)^{1/2}$ where $k_B$ is the Boltzmann constant and $m_H$ is the mass of the atom hydrogen. Considering that the pressure broadening has no significant effect on the spectral line widths in low $n_e$ condition, we can derive the microturbulent broadening $\Delta V_{\rm tur} = (\Delta V^2 - \Delta V_{\rm th}^2)^{1/2}$. The right panel in Figure 10 shows the correlation between microturbulent broadening $\Delta V_{\rm tur}$ and $R_{\rm eff}$. The

$\Delta V_{\rm tur}$ decreases with increasing $R_{\rm eff}$. The correlation coefficient is -0.22 with p-value $\ll 0.001$. $\Delta V_{\rm tur}$ is weakly correlated with $R_{\rm eff}$. It seems that dynamical mechanisms could contribute more to the line broadening for smaller ($R_{\rm eff} < 0.1$ pc) UC HII regions.

### 4.3 Variance of electron temperature in the Galactic Plane

As shown in the middle panels of Figure 3, some UC HII regions show low electron temperature below 6000 K. Shaver et al. (1983) also found that some HII regions (the observations from Parkes 64-m radio telescope) have low electron temperatures below 5000 K, and argued that the low electron temperature is consistent with the discovery of hydrogen recombination lines with narrow linewidths (narrower than 15 km s$^{-1}$). Paladini et al. (2004) also found some HII regions have electron temperatures around 2000 to 5000 K.

Shaver et al. (1983) found that there is a linear relationship between electron temperatures ($T_e$) of HII regions and their galactocentric distances ($R_{\rm GC}$). Afflerbach et al. (1996) analyzed 17 UC HII regions distributed from 4 to 11 kpc in $R_{\rm GC}$ using a non-LTE model and found a similar correlation. Paladini et al. (2004) summarized all the electron temperatures of HII regions in literatures, and confirm the correlation between $T_e$ and $R_{\rm GC}$, which is:

$$T_e = (4166 \pm 124) + (314 \pm 20) R_{\rm GC}. \quad (7)$$

The grey dots in left panel of Figure 11 shows the data from Paladini et al. (2004). The solid grey line shows the least-squares linear relationship that found by Paladini et al. (2004). The dotted dashed line shows the correlation found by Shaver et al. (1983). Balser et al. (2011) also found similar results. The grey cross dots and dashed line show the data and linear fit from Balser et al. (2011). The data from Afflerbach et al. (1996) are shown as blue triangles in the right panel of Figure 11. The yellow dots in Figure 11 represent our data for the UC HII regions in the ATOMS sources. The linear fitting of the data in our sample is:

$$T_e = (6271 \pm 481) + (135 \pm 83) R_{\rm GC}. \quad (8)$$

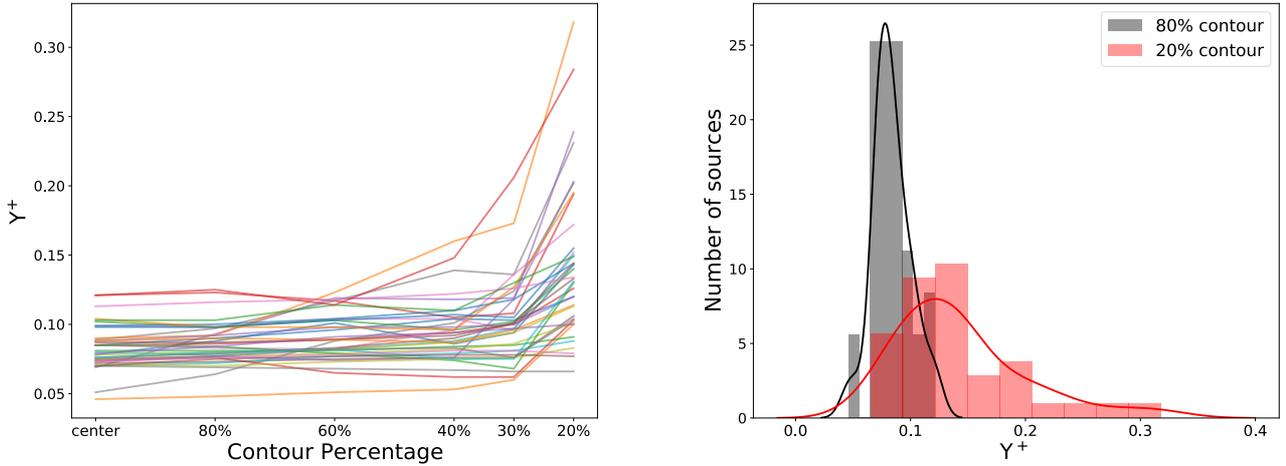

**Figure 7.** Left panel: Y$^+$ in different contour regions. Different colors represent different sources. Right panel: The distribution of Y$^+$ within 80% (black) and 20% contours (red).

The correlation coefficient is 0.17 with p-value ≪ 0.001. However, in contrast to previous works, we do not witness a significant increasing trend in the correlation between T$_e$ and $R_{GC}$ in our data. This could be due to the small dynamical range of $R_{GC}$ in our sample. The $R_{GC}$ for the majority of UC H II regions in our sample is around 4-8 kpc.

However, the T$_e$ at certain $R_{GC}$ in our sample is systematically higher than that in previous works (Paladini et al. 2004; Balser et al. 2011). We noticed that the data compiled by Paladini et al. (2004) and Balser et al. (2011) were mainly collected from single dish observations and their sample only contain low density H II regions. However, the H II regions in our sample are more compact, probably having a higher collision de-excitation rate of the metal lines and a lower cooling efficiency, which could result in a higher T$_e$ (Draine 2011). While the data compiled by Afflerbach et al. (1996) was collected by the Very Large Array. The fundamental limitation of our and their study are that the range in $R_{GC}$ is narrow and their samples are small. One could greatly strengthen the results presented here by increasing the $R_{GC}$ range and sample size. It is especially important to determine if the T$_e$ gradient flattens for $R_{GC}$ > 8 kpc.

## 5 CONCLUSION

In this paper, we have studied 146 ATOMS sources using RRLs H40$\alpha$ and He40$\alpha$ as well as the 3 mm continuum emission. We derived basic parameters for these UC H II regions. Our main results are summarized as follows:

(i) We extracted 94 and 44 compact sources from H40$\alpha$ and He40$\alpha$, respectively. These compact sources are resolved UC H II regions.

(ii) Within UC H II regions, the observed ionic abundance ratio Y$^+$ seems to be higher in the boundary region than in the center. The reason of this phenomenon should be further studied through future higher sensitivity and higher resolution observations with more transitions of RRLs.

(iii) Pressure broadening has no significant effect on the line widths of H40$\alpha$. The microturbulent broadening decreasing with increasing R$_{eff}$ indicate that dynamical mechanisms may contribute more to the line broadening for smaller (R$_{eff}$ <0.1 pc) UC H II regions.

(iv) The electron temperatures of UC H II regions in our sample seem to weakly depend on Galactocentric distance (R$_{GC}$).

## ACKNOWLEDGEMENTS

This paper makes use of the following ALMA data: ADS/JAO.ALMA#2019.1.00685.S. ALMA is a partnership of ESO (representing its member states), NSF (USA), and NINS (Japan), together with NRC (Canada), MOST and ASIAA (Taiwan), and KASI (Republic of Korea), in cooperation with the Republic of Chile. The Joint ALMA Observatory is operated by ESO, AUI/NRAO, and NAOJ.

This work has been supported by the National Key R&D Program of China (No. 2022YFA1603100). Tie Liu acknowledges the supports by National Natural Science Foundation of China (NSFC) through grants No.12073061 and No.12122307, the international partnership program of Chinese Academy of Sciences through grant No.114231KYSB20200009, Shanghai Pujiang Program 20PJ1415500 and the science research grants from the China Manned Space Project with no. CMS-CSST-2021-B06. Fengyao Zhu acknowledges the supports by National Natural Science Foundation of China (NSFC) through grants No.12003055 and Key Research Project of Zhejiang Lab (No. 2021PE0AC03). H.-L. Liu is supported by National Natural Science Foundation of China (NSFC) through the grant No.12103045. G.G. and L.B. gratefully acknowledge support by the ANID BASAL projects ACE210002 and FB210003. K.T. was supported by JSPS KAKENHI (Grant Number 20H05645). L.B. gratefully acknowledges support by the ANID BASAL projects ACE210002 and FB210003

## DATA AVAILABILITY

The rawdata are available in ALMA archive. The derived data underlying this article are available in the article and in its online supplementary material.

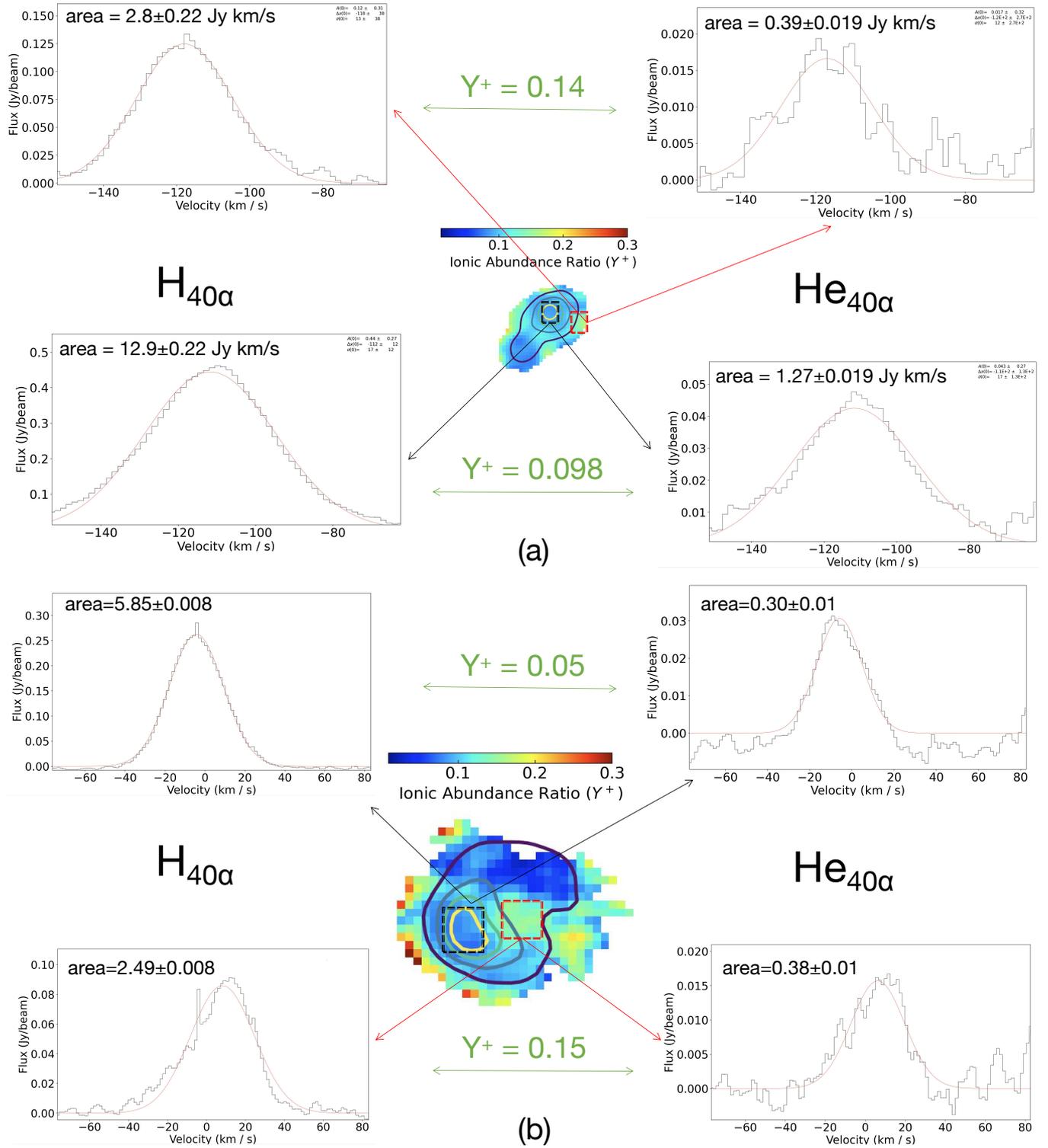

**Figure 8.** (a) The central panel is the Y$^+$ map of source I15567-5236. (b) The central panel is the Y$^+$ map of source I09002-4732. The black contour represents the intensity of H40$\alpha$. Countours are from 20% to 80% in step of 20% of peak value. The red and black box indicate the boundary and central area of source, respectively. The spectrums in the left and right show the H40$\alpha$ and He40$\alpha$. The Y$^+$ of the boundary and centre are 0.14 and 0.098.

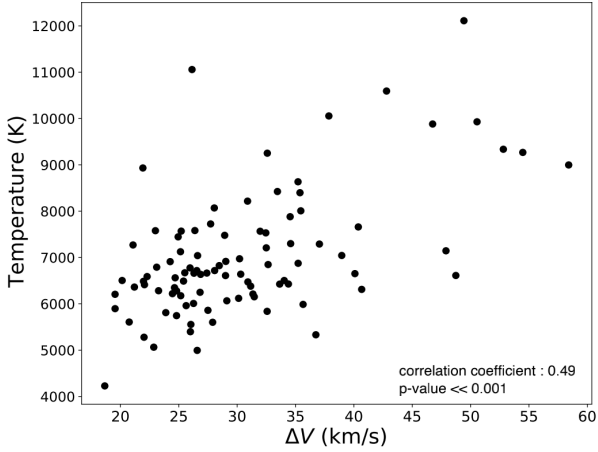

**Figure 9.** The correlation between line width and temperature.

## REFERENCES

Afflerbach A., Churchwell E., Acord J. M., Hofner P., Kurtz S., Depree C. G., 1996, ApJS, 106, 423
Balser D. S., Rood R. T., Bania T. M., Anderson L. D., 2011, ApJ, 738, 27
Brocklehurst M., 1971, MNRAS, 153, 471
Brocklehurst M., Seaton M. J., 1972, MNRAS, 157, 179
Bronfman L., Nyman L. A., May J., 1996, A&AS, 115, 81
Caswell J. L., Haynes R. F., 1987, A&A, 171, 261
Churchwell E., 2002, ARA&A, 40, 27
Churchwell E., Smith L. F., Mathis J., Mezger P. G., Huchtmeier W., 1978, A&A, 70, 719
Churchwell E., Sievers A., Thum C., 2010, A&A, 513, A9
Downes D., Wilson T. L., Bieging J., Wink J., 1980, A&AS, 40, 379
Draine B. T., 2011, Physics of the Interstellar and Intergalactic Medium
Garay G., Lizano S., 1999, PASP, 111, 1049
Gordon M. A., Sorochenko R. L., 2002, Radio Recombination Lines. Their Physics and Astronomical Applications. Vol. 282, doi:10.1007/978-0-387-09604-9,
Habing H. J., Israel F. P., 1979, ARA&A, 17, 345
Keto E., Zhang Q., Kurtz S., 2008, ApJ, 672, 423
Kim W. J., Wyrowski F., Urquhart J. S., Menten K. M., Csengeri T., 2017, A&A, 602, A37
Kim W. J., Urquhart J. S., Wyrowski F., Menten K. M., Csengeri T., 2018, A&A, 616, A107
Krumholz M. R., et al., 2014, in Beuther H., Klessen R. S., Dullemond C. P., Henning T., eds, Protostars and Planets VI. p. 243 (arXiv:1401.2473), doi:10.2458/azu_uapress_9780816531240-ch011
Liu T., et al., 2016, ApJ, 829, 59
Liu T., et al., 2020a, MNRAS, 496, 2790
Liu T., et al., 2020b, MNRAS, 496, 2821
Liu H.-L., et al., 2021, MNRAS, 505, 2801
Liu H.-L., et al., 2022, MNRAS, 511, 501
Luisi M., Anderson L. D., Balser D. S., Bania T. M., Wenger T. V., 2016, ApJ, 824, 125
Nguyen-Luong Q., et al., 2017, ApJ, 844, L25
Paladini R., Davies R. D., De Zotti G., 2004, MNRAS, 347, 237
Peters T., Longmore S. N., Dullemond C. P., 2012, MNRAS, 425, 2352
Pratap P., Snyder L. E., Batrla W., 1992, ApJ, 387, 241
Quireza C., Rood R. T., Bania T. M., Balser D. S., Maciel W. J., 2006, ApJ, 653, 1226
Roshi D. A., Plunkett A., Rosero V., Vaddi S., 2012, ApJ, 749, 49
Roshi D. A., Churchwell E., Anderson L. D., 2017, ApJ, 838, 144
Shaver P. A., McGee R. X., Newton L. M., Danks A. C., Pottasch S. R., 1983, MNRAS, 204, 53
Wink J. E., Wilson T. L., Bieging J. H., 1983, A&A, 127, 211
Zhang C., et al., 2022, MNRAS, 510, 4998
Zhou J.-W., et al., 2021, MNRAS, 508, 4639
Zhu F.-Y., Zhu Q.-F., Wang J.-Z., Zhang J.-S., 2019, ApJ, 881, 14
Zhu F. Y., Wang J. Z., Zhu Q. F., Zhang J. S., 2022, A&A, 665, A94
Zinnecker H., Yorke H. W., 2007, ARA&A, 45, 481

## APPENDIX A: DERIVATION OF ELECTRON TEMPERATURE

Assuming that the 3 mm radio continuum emission is mainly from free-free emission, the electron temperature can be calculated from intensity ratios between H40$\alpha$ RRL and 3 mm continuum. The hot plasma in an H II region gives rise to the emission of thermal bremsstrahlung. This free-free radiation causes a continuum opacity at frequency $\nu$ with the Rayleigh-Jeans approximation as below (Gordon & Sorochenko 2002).

$$\kappa_{\nu,c} = 9.77 \times 10^{-3} \frac{N_e N_i}{\nu^2 T^{3/2}} [17.72 + ln\frac{T_e^{3/2}}{\nu}] \quad (A1)$$

where the numerical version of this equation also in CGS units, which gives $\kappa_c$ in units of cm$^{-1}$. The $N_e$ and $N_i$ is the electron and ion number densities, respectively. Likewise, the plasma emits thermal radiation in LTE conditions with an emissivity of

$$j_{\nu,c} = B_\nu(T)\kappa_{\nu,c} \quad (A2)$$

where $B_\nu(T)$ denotes the intensity of a blackbody of temperature T at frequency $\nu$.

For the hydrogen recombination lines from upper state m to low state n. The corresponding line absorption coefficient is

$$\kappa_{\nu,L} = \frac{h\nu}{4\pi}\phi_\nu(N_n B_{n,m} - N_m B_{m,n}) \quad (A3)$$

with the Planck constant h, the Einstein coefficients $B_{n,m}$ and $B_{m,n}$ for absorption and stimulated emission, respectively. The line profile function including the thermal, turbulent and pressure broadenings is provided by $\phi_\nu$ (Peters et al. 2012). And $N_k = b_k N_k^{LTE}$ is the number densities of hydrogen atoms in state k (for k = m, n) with the departure coefficient $b_k$ (Zhu et al. 2019, 2022). The line emissivity is

$$j_{\nu,L} = \frac{h\nu}{4\pi}\phi_\nu N_m A_{m,n} \quad (A4)$$

$A_{m,n}$ is the Einstein coefficient for spontaneous emission. The detail method of calculating $A_{m,n}$ and $B_{n,m}$ is given by Brocklehurst (1971). We assume an H II region in our sample to be a homogeneous medium of thickness D = 2R. The total optical depth is given by

$$\tau_\nu = \tau_{\nu,c} + \tau_{\nu,L} = (\kappa_{\nu,c} + \kappa_{\nu,L})D \quad (A5)$$

Then, the source function is written as below (Peters et al. 2012)

$$S_\nu = \frac{j_{\nu,c} + j_{\nu,L}}{\kappa_{\nu,c} + \kappa_{\nu,L}} \quad . \quad (A6)$$

The total intensity at frequency $\nu$ is

$$I_\nu = S_\nu(1 - e^{-\tau_\nu}) \quad . \quad (A7)$$

The continuum intensity $I_{\nu,L}$ and the line intensity $I_{\nu,c}$ are calculated as below

$$I_{\nu,C} = B_\nu(T)(1 - e^{-\tau_{\nu,c}}) \quad , \quad (A8)$$

and

$$I_{\nu,L} = S_\nu(1 - e^{-\tau_\nu}) - B_\nu(T)(1 - e^{-\tau_{\nu,c}}) \quad . \quad (A9)$$

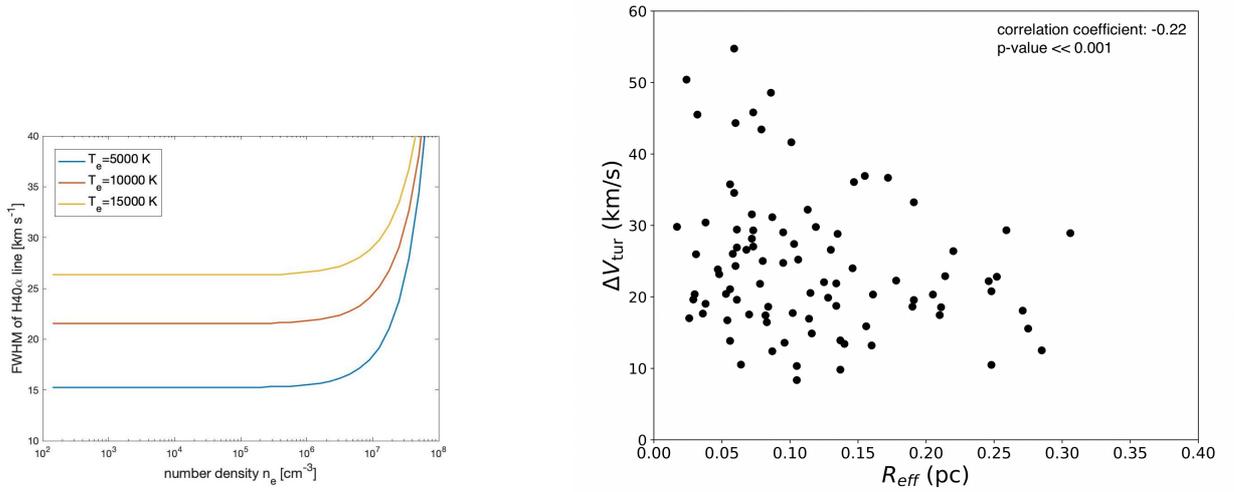

**Figure 10.** Left panel: the effect of thermal and pressure broadening on the spectral line widths at different $n_e$ and temperature. Right panel: the correlation between $\Delta V_{\rm tur}$ and $R_{\rm eff}$.

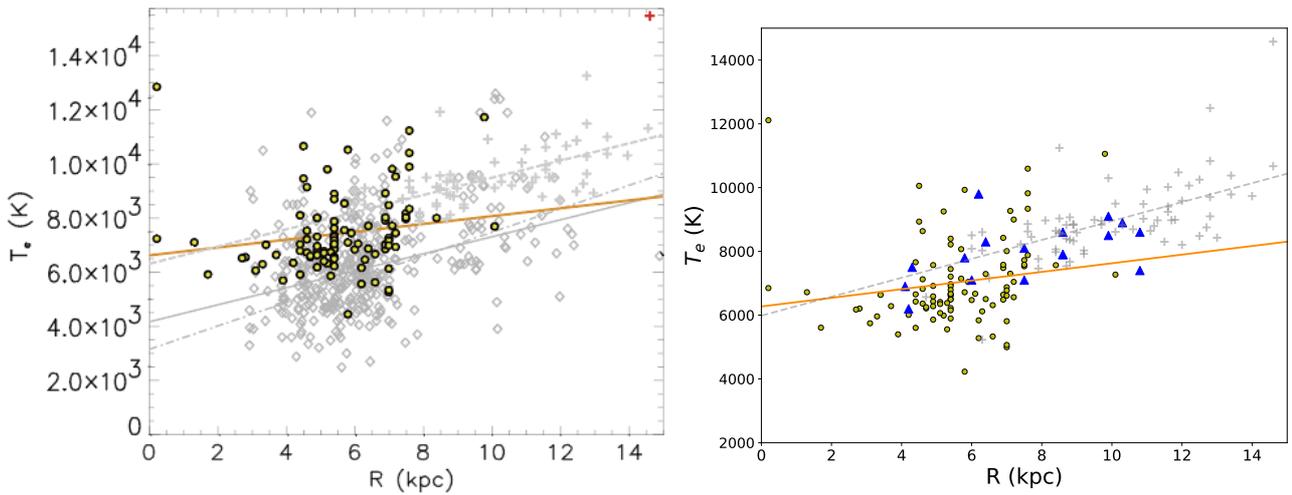

**Figure 11.** Electron temperatures versus $R_{\rm GC}$. The yellow dots are the samples of our paper. The red solid line shows the linear fit to the data. The gray cross dots and dashed line are the data from Balser et al. (2011). Left: The grey dots are the data from Paladini et al. (2004). The solid grey line shows least-squares linear relationship (Equation 8) that found by Paladini et al. (2004). The dot dashed line that found by Shaver et al. (1983). Right: The blue triangles are the data from Afflerbach et al. (1996).

In our sample of H II regions, there is no evidence for strong maser of hydrogen recombination line. So the observed line profile should be similar to the line profile function. Because the line and continuum intensities are functions of $T_e$, $n_e$, $\phi_\nu$ and D, the values of electron temperature and density can be estimated from the observed thermal-emission line and continuum intensities when the thickness of an H II region and the line profile function are known.

The method of estimating $T_e$ and $n_e$ in H II regions is similar with that in Zhu et al. (2022). However, the observable parameters are simulated values in that work, but are measured values in this work. First, the continuum and frequency-integrated H40$\alpha$ line intensities are measured in the observations. Second, a group of two random numbers are generated from the normal distribution based on the two measured intensities. Third, the simulated line and continuum intensities are calculated for the given values of $T_e$ and $n_e$ by using Eq. A8 and A9. $Y^+$ is used to estimate the number ratio of He$^+$ to H$^+$. We adjust the values of $T_e$ and $n_e$ until the simulated values fit the random values generated from the normal distribution. The least-squares method is used in the fitting process. The best fit values of $T_e$ and $n_e$ are the estimates. After a large number of groups of two random numbers are generated and fitted, the distributions of the estimated $T_e$ and $n_e$ are created. The mean values of the distributions are written as the estimates of $T_e$ and $n_e$.

## APPENDIX B: DERIVATION OF ELECTRON DENSITY USING 3MM CONTINUUM

Electron density can also be estimated from the 3 mm continuum emission. The intensity of continuum emission is computed by

$$I_{\nu,C} = B_\nu(T)\, \tau_{\nu,c} = B_\nu(T)\, \kappa_{\nu,c} D \quad (B1)$$

All quantities are in cgs units. Combining Equation B1 and A1 that assume a Rayleigh-Jeans approximation, we get

$$I_{\nu,C} = 2.931 \times 10^{-39} T^{-0.5} [17.72 + \ln \frac{T_e^{3/2}}{\nu}] EM \quad (B2)$$

where $EM = N_e N_i D$ and the numerical version of this equation also in CGS units. Then switching units, we get

$$EM = \frac{I_{\nu,c}}{879.3\, T^{-0.5}[17.72 + \ln \frac{T_e^{3/2}}{\nu}]} \quad (B3)$$

where EM in unit $cm^{-6}$ pc, $I_{\nu,c}$ in unit Jy, $\nu$ in unit Hz After getting the value of EM, we use $n_e = \sqrt{\frac{EM}{D}}$ to get electron density.

## APPENDIX C: OBVERVATION PARAMETERS

This paper has been typeset from a T<sub>E</sub>X/L<sup>A</sup>T<sub>E</sub>X file prepared by the author.

**Table C1.** The derived parameters of RRLs and 3mm continuum.

| IRAS | ID | Peak Intensity H40$\alpha$ (Jy beam$^{-1}$) | vlsr H40$\alpha$ (km/s) | $\Delta V_{H40\alpha}$ (km/s) | peak intensity He40$\alpha$ (Jy beam$^{-1}$) | vlsr He40$\alpha$ (km/s) | $\Delta V_{He40\alpha}$ (km/s) | Intensity 3mm (Jy) |
|---|---|---|---|---|---|---|---|---|
| I09002-4732 | 1 | 0.132 ± 0.054 | -0.3 ± 1.2 | 32.0 ± 1.2 | 0.012 ± 0.011 | -1.6 ± 1.2 | 32.3 ± 1.2 | 13.37 ± 0.01 |
| I11298-6155 | 1 | 0.012 ± 0.009 | 39.5 ± 1.2 | 21.1 ± 1.2 | - | - | - | 0.57 ± 0.00 |
| I12320-6122 | 1 | 0.155 ± 0.070 | -41.4 ± 1.2 | 39.0 ± 1.2 | 0.006 ± 0.014 | -44.9 ± 1.2 | 44.2 ± 1.2 | 1.92 ± 0.00 |
| I12326-6245 | 1 | 0.425 ± 0.815 | -63.4 ± 1.2 | 58.4 ± 1.2 | 0.032 ± 0.158 | -60.5 ± 1.2 | 72.6 ± 1.2 | 8.26 ± 0.02 |
| I12383-6128 | 1 | 0.026 ± 0.015 | -39.2 ± 1.2 | 24.7 ± 1.2 | - | - | - | 0.41 ± 0.00 |
| I12572-6316 | 1 | 0.013 ± 0.012 | 24.2 ± 1.2 | 26.1 ± 1.2 | - | - | - | 0.20 ± 0.00 |
| I13080-6229 | 1 | 0.055 ± 0.015 | -36.5 ± 1.2 | 28.9 ± 1.2 | 0.007 ± 0.003 | -36.8 ± 1.2 | 25.2 ± 1.2 | 9.12 ± 0.00 |
| I13111-6228 | 1 | 0.009 ± 0.011 | -34.9 ± 1.2 | 33.5 ± 1.2 | - | - | - | 0.21 ± 0.00 |
| I13291-6229 | 1 | 0.015 ± 0.014 | -36.7 ± 1.2 | 26.6 ± 1.2 | - | - | - | 0.42 ± 0.00 |
| I13291-6229 | 2 | 0.011 ± 0.012 | -31.7 ± 1.2 | 23.9 ± 1.2 | - | - | - | 0.35 ± 0.00 |
| I13291-6249 | 1 | 0.046 ± 0.018 | -38.7 ± 1.2 | 34.6 ± 1.2 | 0.005 ± 0.007 | -40.6 ± 1.2 | 22.0 ± 1.2 | 2.27 ± 0.00 |
| I13471-6120* | 1 | 0.335 ± 0.098 | -57.1 ± 1.2 | 34.0 ± 1.2 | 0.023 ± 0.021 | -57.8 ± 1.2 | 36.4 ± 1.2 | 2.90 ± 0.01 |
| I14013-6105 | 1 | 0.109 ± 0.062 | -57.9 ± 1.2 | 37.0 ± 1.2 | 0.010 ± 0.011 | -59.7 ± 1.2 | 34.3 ± 1.2 | 4.71 ± 0.01 |
| I14050-6056 | 1 | 0.039 ± 0.010 | -49.1 ± 1.2 | 40.7 ± 1.2 | 0.003 ± 0.007 | -47.8 ± 1.2 | 43.6 ± 1.2 | 3.50 ± 0.00 |
| I14382-6017 | 1 | 0.011 ± 0.004 | -58.3 ± 1.2 | 30.9 ± 1.2 | - | - | - | 3.09 ± 0.00 |
| I14453-5912 | 1 | 0.005 ± 0.004 | -40.8 ± 1.2 | 36.7 ± 1.2 | - | - | - | 1.29 ± 0.00 |
| I15254-5621* | 1 | 0.364 ± 0.191 | -70.8 ± 1.2 | 47.9 ± 1.2 | 0.027 ± 0.023 | -73.0 ± 1.2 | 54.4 ± 1.2 | 4.69 ± 0.02 |
| I15290-5546 | 1 | 0.148 ± 0.021 | -89.4 ± 1.2 | 25.2 ± 1.2 | 0.013 ± 0.014 | -88.9 ± 1.2 | 19.3 ± 1.2 | 2.26 ± 0.00 |
| I15290-5546 | 2 | 0.155 ± 0.040 | -88.4 ± 1.2 | 29.0 ± 1.2 | 0.013 ± 0.012 | -90.6 ± 1.2 | 32.7 ± 1.2 | 5.16 ± 0.01 |
| I15384-5348 | 1 | 0.029 ± 0.008 | -41.5 ± 1.2 | 22.0 ± 1.2 | 0.004 ± 0.006 | -40.7 ± 1.2 | 22.9 ± 1.2 | 4.87 ± 0.00 |
| I15408-5356 | 1 | 0.027 ± 0.004 | -42.8 ± 1.2 | 23.0 ± 1.2 | 0.002 ± 0.003 | -48.4 ± 1.2 | 51.5 ± 1.2 | 7.80 ± 0.00 |
| I15411-5352 | 1 | 0.076 ± 0.024 | -41.0 ± 1.2 | 30.3 ± 1.2 | 0.007 ± 0.007 | -43.4 ± 1.2 | 28.8 ± 1.2 | 9.75 ± 0.00 |
| I15439-5449 | 1 | 0.058 ± 0.023 | -52.5 ± 1.2 | 26.6 ± 1.2 | - | - | - | 1.20 ± 0.00 |
| I15502-5302 | 1 | 0.953 ± 0.344 | -94.8 ± 1.2 | 35.2 ± 1.2 | 0.082 ± 0.045 | -95.8 ± 1.2 | 33.7 ± 1.2 | 13.67 ± 0.03 |
| I15520-5234 | 1 | 0.300 ± 0.061 | -38.9 ± 1.2 | 32.6 ± 1.2 | - | - | - | 7.08 ± 0.01 |
| I15567-5236 | 1 | 0.230 ± 0.124 | -112.0 ± 1.2 | 40.1 ± 1.2 | 0.021 ± 0.019 | -113.6 ± 1.2 | 39.5 ± 1.2 | 8.88 ± 0.01 |
| I15570-5227 | 1 | 0.010 ± 0.007 | -103.2 ± 1.2 | 33.6 ± 1.2 | - | - | - | 2.11 ± 0.00 |
| I16037-5223* | 1 | 0.138 ± 0.078 | -81.9 ± 1.2 | 34.4 ± 1.2 | 0.010 ± 0.024 | -83.7 ± 1.2 | 36.6 ± 1.2 | 1.44 ± 0.00 |
| I16037-5223* | 2 | 0.094 ± 0.021 | -81.2 ± 1.2 | 22.3 ± 1.2 | 0.011 ± 0.020 | -81.9 ± 1.2 | 15.9 ± 1.2 | 1.01 ± 0.00 |
| I16037-5223* | 3 | 0.043 ± 0.025 | -78.9 ± 1.2 | 27.5 ± 1.2 | 0.002 ± 0.011 | -92.2 ± 1.2 | 71.0 ± 1.2 | 0.40 ± 0.00 |
| I16060-5146 | 1 | 1.439 ± 0.583 | -90.8 ± 1.2 | 37.9 ± 1.2 | 0.112 ± 0.062 | -91.3 ± 1.2 | 39.2 ± 1.2 | 24.95 ± 0.05 |
| I16065-5158 | 1 | 0.030 ± 0.036 | -57.0 ± 1.2 | 32.6 ± 1.2 | - | - | - | 2.88 ± 0.00 |
| I16065-5158 | 2 | 0.017 ± 0.049 | -49.6 ± 1.2 | 35.7 ± 1.2 | - | - | - | 0.75 ± 0.00 |
| I16071-5142 | 1 | 0.032 ± 0.015 | -82.1 ± 1.2 | 21.9 ± 1.2 | - | - | - | 1.15 ± 0.00 |
| I16132-5039 | 1 | 0.008 ± 0.007 | -48.3 ± 1.2 | 18.7 ± 1.2 | - | - | - | 0.25 ± 0.00 |
| I16158-5055 | 1 | 0.015 ± 0.010 | -51.7 ± 1.2 | 20.1 ± 1.2 | - | - | - | 0.71 ± 0.00 |
| I16164-5046 | 1 | 1.093 ± 0.529 | -65.0 ± 1.2 | 30.2 ± 1.2 | 0.082 ± 0.073 | -67.1 ± 1.2 | 29.4 ± 1.2 | 18.53 ± 0.03 |
| I16172-5028 | 1 | 0.798 ± 0.241 | -53.3 ± 1.2 | 35.4 ± 1.2 | 0.079 ± 0.042 | -52.1 ± 1.2 | 37.2 ± 1.2 | 12.87 ± 0.03 |
| I16177-5018 | 1 | 0.090 ± 0.114 | -55.7 ± 1.2 | 29.0 ± 1.2 | 0.010 ± 0.036 | -55.8 ± 1.2 | 35.4 ± 1.2 | 2.19 ± 0.00 |
| I16177-5018 | 2 | 0.124 ± 0.043 | -49.4 ± 1.2 | 24.3 ± 1.2 | 0.014 ± 0.034 | -49.9 ± 1.2 | 22.9 ± 1.2 | 5.93 ± 0.00 |
| I16177-5018 | 3 | 0.089 ± 0.059 | -51.6 ± 1.2 | 26.3 ± 1.2 | 0.008 ± 0.030 | -50.2 ± 1.2 | 28.1 ± 1.2 | 0.83 ± 0.00 |
| I16177-5018 | 4 | 0.043 ± 0.024 | -48.5 ± 1.2 | 25.4 ± 1.2 | 0.005 ± 0.018 | -48.8 ± 1.2 | 34.1 ± 1.2 | 2.30 ± 0.00 |
| I16177-5018 | 5 | 0.076 ± 0.015 | -50.1 ± 1.2 | 23.1 ± 1.2 | 0.009 ± 0.015 | -50.0 ± 1.2 | 22.0 ± 1.2 | 7.18 ± 0.00 |
| I16297-4757 | 1 | 0.010 ± 0.007 | -79.9 ± 1.2 | 26.3 ± 1.2 | - | - | - | 1.15 ± 0.00 |
| I16304-4710 | 1 | 0.022 ± 0.011 | -60.6 ± 1.2 | 24.8 ± 1.2 | - | - | - | 0.74 ± 0.00 |
| I16313-4729* | 1 | 0.023 ± 0.026 | -78.6 ± 1.2 | 40.4 ± 1.2 | - | - | - | 0.34 ± 0.00 |
| I16318-4724 | 1 | 0.011 ± 0.012 | -111.4 ± 1.2 | 25.6 ± 1.2 | - | - | - | 0.50 ± 0.00 |
| I16330-4725 | 1 | 0.088 ± 0.036 | -75.5 ± 1.2 | 28.5 ± 1.2 | - | - | - | 1.60 ± 0.00 |
| I16330-4725 | 2 | 0.032 ± 0.024 | -78.3 ± 1.2 | 25.1 ± 1.2 | - | - | - | 1.27 ± 0.00 |
| I16348-4654* | 1 | 0.158 ± 0.036 | -48.6 ± 1.2 | 30.9 ± 1.2 | - | - | - | 1.66 ± 0.01 |
| I16351-4722* | 1 | 0.070 ± 0.030 | -31.6 ± 1.2 | 28.0 ± 1.2 | - | - | - | 0.88 ± 0.00 |
| I16385-4619 | 1 | 0.048 ± 0.010 | -113.3 ± 1.2 | 24.8 ± 1.2 | - | - | - | 0.78 ± 0.00 |
| I16445-4459 | 1 | 0.030 ± 0.011 | -128.1 ± 1.2 | 19.5 ± 1.2 | - | - | - | 0.49 ± 0.00 |
| I16458-4512* | 1 | 0.089 ± 0.020 | -54.6 ± 1.2 | 29.1 ± 1.2 | - | - | - | 0.71 ± 0.00 |
| I16506-4512 | 1 | 0.013 ± 0.002 | -27.9 ± 1.2 | 25.5 ± 1.2 | - | - | - | 5.85 ± 0.00 |

**Table C1** – *continued*

| IRAS | ID | Peak Intensity H40α (Jy beam$^{-1}$) | vlsr H40α (km/s) | $\Delta V_{H40\alpha}$ (km/s) | peak intensity He40α (Jy beam$^{-1}$) | vlsr He40α (km/s) | $\Delta V_{He40\alpha}$ (km/s) | Intensity 3mm (Jy) |
|---|---|---|---|---|---|---|---|---|
| I17006-4215 | 1 | 0.113 ± 0.019 | -24.1 ± 1.2 | 30.1 ± 1.2 | 0.007 ± 0.008 | -27.1 ± 1.2 | 38.7 ± 1.2 | 3.46 ± 0.00 |
| I17016-4124* | 1 | 0.166 ± 0.058 | -32.2 ± 1.2 | 35.5 ± 1.2 | - | - | - | 2.10 ± 0.01 |
| I17136-3617 | 1 | 0.123 ± 0.073 | -6.6 ± 1.2 | 35.2 ± 1.2 | 0.009 ± 0.012 | -5.4 ± 1.2 | 36.5 ± 1.2 | 7.83 ± 0.00 |
| I17143-3700 | 1 | 0.050 ± 0.014 | -34.6 ± 1.2 | 31.4 ± 1.2 | 0.001 ± 0.011 | -40.0 ± 1.2 | 37.1 ± 1.2 | 0.76 ± 0.00 |
| I17160-3707 | 1 | 0.029 ± 0.004 | -72.2 ± 1.2 | 25.2 ± 1.2 | - | - | - | 4.06 ± 0.00 |
| I17175-3544 | 1 | 0.394 ± 0.103 | -4.5 ± 1.2 | 26.0 ± 1.2 | - | - | - | 12.71 ± 0.01 |
| I17204-3636 | 1 | 0.023 ± 0.020 | -5.5 ± 1.2 | 22.1 ± 1.2 | - | - | - | 0.51 ± 0.00 |
| I17220-3609 | 1 | 0.325 ± 0.041 | -92.8 ± 1.2 | 26.5 ± 1.2 | 0.022 ± 0.017 | -93.6 ± 1.2 | 21.5 ± 1.2 | 3.68 ± 0.01 |
| I17233-3606 | 1 | 0.069 ± 0.012 | -1.8 ± 1.2 | 22.9 ± 1.2 | - | - | - | 0.98 ± 0.00 |
| I17244-3536 | 1 | 0.020 ± 0.010 | -12.5 ± 1.2 | 48.7 ± 1.2 | - | - | - | 0.62 ± 0.00 |
| I17258-3637 | 1 | 0.410 ± 0.047 | -12.6 ± 1.2 | 28.1 ± 1.2 | 0.048 ± 0.011 | -13.3 ± 1.2 | 25.3 ± 1.2 | 42.49 ± 0.01 |
| I17271-3439 | 1 | 0.109 ± 0.028 | -13.8 ± 1.2 | 31.2 ± 1.2 | - | - | - | 2.15 ± 0.00 |
| I17441-2822* | 1 | 0.889 ± 1.443 | 62.3 ± 1.2 | 49.4 ± 1.2 | 0.074 ± 0.151 | 62.5 ± 1.2 | 51.7 ± 1.2 | 25.06 ± 0.04 |
| I17441-2822* | 2 | 0.507 ± 0.626 | 59.7 ± 1.2 | 32.7 ± 1.2 | 0.039 ± 0.110 | 59.4 ± 1.2 | 32.9 ± 1.2 | 13.41 ± 0.01 |
| I17455-2800 | 1 | 0.014 ± 0.003 | -20.9 ± 1.2 | 20.8 ± 1.2 | 0.003 ± 0.003 | -22.8 ± 1.2 | 18.8 ± 1.2 | 2.95 ± 0.00 |
| I17545-2357* | 1 | 0.079 ± 0.012 | 7.7 ± 1.2 | 24.5 ± 1.2 | - | - | - | 0.49 ± 0.00 |
| I17599-2148 | 1 | 0.012 ± 0.022 | 28.2 ± 1.2 | 24.9 ± 1.2 | - | - | - | 0.63 ± 0.00 |
| I17599-2148 | 2 | 0.023 ± 0.024 | 21.6 ± 1.2 | 32.5 ± 1.2 | - | - | - | 1.34 ± 0.00 |
| I17599-2148 | 3 | 0.026 ± 0.014 | 21.1 ± 1.2 | 27.4 ± 1.2 | - | - | - | 1.00 ± 0.00 |
| I18032-2032* | 1 | 0.054 ± 0.019 | 11.5 ± 1.2 | 26.9 ± 1.2 | - | - | - | 0.49 ± 0.00 |
| I18110-1854 | 1 | 0.148 ± 0.015 | 41.7 ± 1.2 | 24.6 ± 1.2 | 0.008 ± 0.006 | 42.3 ± 1.2 | 25.1 ± 1.2 | 3.58 ± 0.00 |
| I18116-1646 | 1 | 0.019 ± 0.003 | 53.0 ± 1.2 | 21.2 ± 1.2 | - | - | - | 6.16 ± 0.00 |
| I18139-1842 | 1 | 0.020 ± 0.007 | 36.2 ± 1.2 | 19.5 ± 1.2 | - | - | - | 0.42 ± 0.00 |
| I18228-1312 | 1 | 0.021 ± 0.020 | 29.8 ± 1.2 | 31.5 ± 1.2 | - | - | - | 1.04 ± 0.00 |
| I18311-0809 | 1 | 0.022 ± 0.008 | 108.6 ± 1.2 | 23.3 ± 1.2 | - | - | - | 0.85 ± 0.00 |
| I18314-0720 | 1 | 0.006 ± 0.002 | 105.8 ± 1.2 | 26.0 ± 1.2 | - | - | - | 1.86 ± 0.00 |
| I18317-0757 | 1 | 0.024 ± 0.008 | 78.7 ± 1.2 | 27.9 ± 1.2 | - | - | - | 3.96 ± 0.00 |
| I18434-0242 | 1 | 0.204 ± 0.018 | 99.0 ± 1.2 | 26.8 ± 1.2 | 0.019 ± 0.005 | 97.7 ± 1.2 | 26.7 ± 1.2 | 9.20 ± 0.01 |
| I18479-0005 | 1 | 0.436 ± 0.126 | 13.9 ± 1.2 | 27.7 ± 1.2 | 0.044 ± 0.088 | 13.5 ± 1.2 | 23.0 ± 1.2 | 6.10 ± 0.01 |
| I18479-0005 | 2 | 0.129 ± 0.130 | 19.4 ± 1.2 | 32.5 ± 1.2 | 0.014 ± 0.042 | 19.3 ± 1.2 | 30.1 ± 1.2 | 2.18 ± 0.00 |
| I18479-0005 | 3 | 0.104 ± 0.027 | 14.8 ± 1.2 | 26.4 ± 1.2 | 0.014 ± 0.034 | 14.9 ± 1.2 | 22.7 ± 1.2 | 3.42 ± 0.00 |
| I18507+0110 | 1 | 1.263 ± 0.551 | 49.1 ± 1.2 | 54.5 ± 1.2 | - | - | - | 24.76 ± 0.05 |
| I18530+0215 | 1 | 0.026 ± 0.010 | 80.8 ± 1.2 | 26.0 ± 1.2 | - | - | - | 0.92 ± 0.00 |
| I19078+0901 | 1 | 0.582 ± 0.560 | 10.4 ± 1.2 | 42.8 ± 1.2 | 0.054 ± 0.117 | 10.1 ± 1.2 | 35.0 ± 1.2 | 17.86 ± 0.03 |
| I19078+0901 | 2 | 0.339 ± 0.611 | 12.9 ± 1.2 | 46.8 ± 1.2 | 0.027 ± 0.073 | 13.6 ± 1.2 | 49.4 ± 1.2 | 6.75 ± 0.01 |
| I19078+0901 | 3 | 0.220 ± 0.444 | 13.5 ± 1.2 | 52.8 ± 1.2 | 0.018 ± 0.049 | 24.0 ± 1.2 | 79.1 ± 1.2 | 3.60 ± 0.02 |
| I19078+0901 | 4 | 0.178 ± 0.083 | -2.2 ± 1.2 | 34.5 ± 1.2 | 0.017 ± 0.024 | -2.3 ± 1.2 | 29.7 ± 1.2 | 2.94 ± 0.01 |
| I19095+0930 | 1 | 0.050 ± 0.043 | 64.9 ± 1.2 | 50.6 ± 1.2 | - | - | - | 0.93 ± 0.00 |
| I19097+0847 | 1 | 0.010 ± 0.006 | 64.3 ± 1.2 | 22.0 ± 1.2 | - | - | - | 0.53 ± 0.00 |

The ∗ symbol after the IRAS name indicates that the sources are unresolved. The peak intensity, velocity ($V_{lsr}$) and line widths ($\Delta V$) of H40α and He40α are derived from the Gaussian fitting of the spectra of each line. The uncertainties of $\Delta V$ is derived by considering the velocity resolution and the uncertainties given by gaussian fitting. The uncertainties of the other parameters are derived from gaussian fitting.